\documentclass[iop,numberedappendix,appendixfloats]{emulateapj}
\usepackage{amsmath}
\usepackage[hidelinks]{hyperref}

\newcommand{\n}{\noindent}
\newcommand{\Ni}{$^{56}$Ni}

\newcommand{\numOfLCs}{57}
\newcommand{\NiPValue}{0.76}

\begin{document}
\submitted{}
\title{Type II supernova energetics and comparison of light curves to shock-cooling models}
\keywords{}
\author{Adam Rubin,\altaffilmark{1} 
Avishay Gal-Yam,\altaffilmark{1} 
Annalisa De Cia,\altaffilmark{1} 
Assaf Horesh,\altaffilmark{1} 
Danny Khazov,\altaffilmark{1} 
Eran O. Ofek,\altaffilmark{1} 
S. R. Kulkarni,\altaffilmark{2,3} 
Iair Arcavi,\altaffilmark{4,5} 
Ilan Manulis,\altaffilmark{1} 
Ofer Yaron,\altaffilmark{1} 
Paul Vreeswijk,\altaffilmark{1} 
Mansi M. Kasliwal,\altaffilmark{6} 
Sagi Ben-Ami,\altaffilmark{7} 
Daniel A. Perley,\altaffilmark{2,8} 
Yi Cao,\altaffilmark{2} 
S. Bradley Cenko,\altaffilmark{9,10} 
Umaa D. Rebbapragada,\altaffilmark{11} 
P. R. Wo\'zniak,\altaffilmark{12} 
Alexei V. Filippenko,\altaffilmark{13} 
K. I. Clubb,\altaffilmark{13} 
Peter E. Nugent,\altaffilmark{13,14} 
Y.-C. Pan,\altaffilmark{15} 
C. Badenes,\altaffilmark{16} 
D. Andrew Howell,\altaffilmark{4,17} 
Stefano Valenti,\altaffilmark{4} 
David Sand,\altaffilmark{18} 
J. Sollerman,\altaffilmark{19} 
Joel Johansson,\altaffilmark{20} 
Douglas C. Leonard,\altaffilmark{21} 
J. Chuck Horst,\altaffilmark{21} 
Stephen F. Armen,\altaffilmark{21} 
Joseph M. Fedrow,\altaffilmark{21,22} 
Robert M. Quimby,\altaffilmark{21,23} 
Paulo Mazzali,\altaffilmark{24,25} 
Elena Pian,\altaffilmark{26,27} 
Assaf Sternberg,\altaffilmark{25,28} 
Thomas Matheson,\altaffilmark{29} 
M. Sullivan,\altaffilmark{30} 
K. Maguire,\altaffilmark{31} 
 and Sanja Lazarevic\altaffilmark{32}}

\altaffiltext{1}{Department of Particle Physics and Astrophysics, Weizmann Institute of Science, 234 Herzl St., Rehovot, Israel; \email{adam.rubin@weizmann.ac.il}}
\altaffiltext{2}{Astronomy Department, California Institute of Technology, Pasadena, CA 91125, USA}
\altaffiltext{3}{Caltech Optical Observatories, California Institute of Technology, Pasadena, CA 91125, USA}
\altaffiltext{4}{Las Cumbres Observatory Global Telescope Network, 6740 Cortona Dr., Suite 102, Goleta, CA 93117, USA}
\altaffiltext{5}{Kavli Institute for Theoretical Physics, University of California, Santa Barbara, CA 93106, USA}
\altaffiltext{6}{Observatories of the Carnegie Institution for Science, 813 Santa Barbara Street, Pasadena, CA 91101, USA}
\altaffiltext{7}{Smithsonian Astrophysical Observatory, Harvard-Smithsonian Center for Astrophysics, 60 Garden St., Cambridge, MA 02138, USA}
\altaffiltext{8}{Dark Cosmology Centre, Niels Bohr Institute, University of Copenhagen, Juliane Maries Vej 30, DK-2100 K{\o}benhavn \O, Denmark}
\altaffiltext{9}{Astrophysics Science Division, NASA Goddard Space Flight Center, Mail Code 661, Greenbelt, MD 20771, USA}
\altaffiltext{10}{Joint Space Science Institute, University of Maryland, College Park, MD 20742, USA}
\altaffiltext{11}{Jet Propulsion Laboratory, California Institute of Technology, Pasadena, CA 91109, USA}
\altaffiltext{12}{Los Alamos National Laboratory, Los Alamos, NM 87545, USA}
\altaffiltext{13}{Department of Astronomy, University of California, Berkeley, CA 94720-3411, USA}
\altaffiltext{14}{Lawrence Berkeley National Laboratory, Berkeley, CA 94720, USA}
\altaffiltext{15}{Astronomy Department, University of Illinois at Urbana-Champaign, 1002 W. Green Street, Urbana, IL 61801, USA}
\altaffiltext{16}{Department of Physics and Astronomy, and Pittsburgh Particle Physics, Astrophysics, and Cosmology Center (PITT-PACC), University of Pittsburgh, 3941 O'Hara St., Pittsburgh, PA 15260, USA}
\altaffiltext{17}{Department of Physics, University of California, Santa Barbara, Broida Hall, Mail Code 9530, Santa Barbara, CA 93106-9530, USA}
\altaffiltext{18}{Physics Department, Texas Tech University, Lubbock, TX 79409, USA}
\altaffiltext{19}{The Oskar Klein Centre, Department of Astronomy, Stockholm University, SE-106 91 Stockholm, Sweden}
\altaffiltext{20}{The Oskar Klein Centre, Department of Physics, Stockholm University,  SE-106 91 Stockholm, Sweden}
\altaffiltext{21}{Department of Astronomy, San Diego State University, San Diego, CA 92182-1221, USA}
\altaffiltext{22}{Yukawa Institute for Theoretical Physics, Kyoto University, Kyoto 606-8502, Japan}
\altaffiltext{23}{Kavli IPMU (WPI), UTIAS, The University of Tokyo, Kashiwa, Chiba, 277-8583, Japan}
\altaffiltext{24}{Astrophysics Research Institute, Liverpool John Moores University, IC2, Liverpool Science Park, 146 Browlow Hill, Liverpool L3 5RF, UK}
\altaffiltext{25}{Max-Planck-Institut f\"ur Astrophysik, Karl-Schwarzschild-Strasse 1, D-85748 Garching, Germany}
\altaffiltext{26}{INAF, Institute of Space Astrophysics and Cosmic Physics, via P. Gobetti 101, 40129 Bologna, Italy}
\altaffiltext{27}{Scuola Normale Superiore, Piazza dei Cavalieri 7, I-56126 Pisa, Italy}
\altaffiltext{28}{Excellence Cluster Universe, Technische Universitat M\"unchen, Boltzmannstr. 2, D-85748 Garching, Germany}
\altaffiltext{29}{National Optical Astronomy Observatory, 950 N. Cherry Avenue, Tucson, AZ 85719, USA}
\altaffiltext{30}{School of Physics and Astronomy, University of Southampton, Southampton SO17 1BJ, UK}
\altaffiltext{31}{ESO, Karl-Schwarzschild-Str. 2, D-85748 Garching, Germany}
\altaffiltext{32}{Department of Astronomy, Faculty of Mathematics, University of Belgrade, Serbia}

\begin{abstract}

	During the first few days after explosion, Type II supernovae (SNe) are dominated by relatively simple physics. Theoretical predictions regarding early-time SN light curves in the ultraviolet (UV) and optical bands are thus quite robust. We present, for the first time, a sample of \numOfLCs{} $R$-band Type II SN light curves that are well monitored during their rise, having $>5$ detections during the first 10 days after discovery, and a well-constrained time of explosion to within 1--3 days. We show that the energy per unit mass ($E/M$) can be deduced to roughly a factor of five by comparing early-time optical data to the model of \cite{rabinak_early_2011}, while the progenitor radius cannot be determined based on $R$-band data alone. We find that Type II SN explosion energies span a range of $E/M=(0.2-20)\times 10^{51}$ erg/(10 M$_\odot$), and have a mean energy per unit mass of $\left\langle E/M \right\rangle = 0.85\times 10^{51}$ erg/(10 M$_\odot$), corrected for Malmquist bias. Assuming a small spread in progenitor masses, this indicates a large intrinsic diversity in explosion energy. Moreover, $E/M$ is positively correlated with the amount of $^{56}$Ni produced in the explosion, as predicted by some recent models of core-collapse SNe. We further present several empirical correlations. The peak magnitude is correlated with the decline rate ($\Delta m_{15}$), the decline rate is weakly correlated with the rise time, and the rise time is not significantly correlated with the peak magnitude. Faster declining SNe are more luminous and have longer rise times. This limits the possible power sources for such events.

\end{abstract}

\maketitle

\section{Introduction}

Despite the recent availability of large samples of Type II SN light curves \citep[e.g.,][]{arcavi_caltech_2012, anderson_characterizing_2014,faran_photometric_2014,faran_sample_2014,sanders_toward_2015,gonzalez-gaitan_rise-time_2015}, there is little high-quality data in the literature against which to test predictions \citep[e.g.,][NS10, RW11]{nakar_early_2010,rabinak_early_2011} regarding early-time light-curve behavior in the ultraviolet (UV) and optical bands. \cite{rabinak_early_2011} showed that it is possible to deduce the progenitor star radius ($R_*$) and energy per unit mass ($E/M$) from the early UV light curve. This is because at early times (in the first 3--4 days after explosion), the light curve is dominated by shock cooling; the photosphere is at the outer edge of the ejecta, and no recombination has set in. RW11 models describe well the handful of available early UV SN light curves \citep{soderberg_extremely_2008,gezari_probing_2008, schawinski_supernova_2008}, and can fit the rate of UV detections by a GALEX/PTF survey \citep{ganot_detection_2014}. 

Recently, \cite{gall_comparative_2015} and \cite{gonzalez-gaitan_rise-time_2015} compared large samples of SN~II light curves to RW11/NS10 shock-cooling models. Both papers compared SN rise times to rise times derived from shock-cooling models: \cite{gall_comparative_2015} used $r$-band data, while \cite{gonzalez-gaitan_rise-time_2015} compared multi-band photometry. Both papers concluded that only models with small radii are consistent with the data --- a conclusion that is in tension with the known association of red supergiants (RSGs) with Type II-P SNe \citep{smartt_death_2009}. However, as we show in Section \ref{sec:discussion}, comparing to models based on their rise time requires the application of the models beyond their validity and leads to rejection of models with larger radii that fit the early-time data well. \cite{valenti_first_2014} and \cite{bose_sn_2015} compared multi-band photometry of SN 2013ej and SN 2013ab to RW11 models, but limited their analysis to the first week after explosion. They found their data to be consistent with RW11 models with radii of 400--600 \; R$_\odot$ and 450--1500 \; R$_\odot$, respectively. 

Basic empirical relations involving the time scales of the rising light curve have yet to be established. This is due to the fact that most of the published SN photometry begins shortly prior to the peak (if at all); light curves that are well sampled during the first days after explosion are still rare. Based on three such events, \cite{gal-yam_real-time_2011} suggested that there may be a trend where more luminous SNe~II-P also rise more slowly. More recently, \cite{faran_photometric_2014} suggested that the rise time and luminosity are uncorrelated, but did not perform a quantitative analysis owing to their small sample size. \cite{gall_comparative_2015} studied the rise times of 19 well-monitored SNe, and concluded that there is a qualitative trend between rise time and peak magnitude, with brighter events having longer rise times. Here we use a sample of \numOfLCs{} spectroscopically confirmed SN~II $R$-band light curves that were well monitored during their rise to test and establish such correlations, and we quantitatively compare 33 of these to shock-cooling models.

\vfill\break
\section{The sample}
Our sample consists of \numOfLCs{} SNe from the Palomar Transient Factory \citep[PTF;][]{law_palomar_2009,rau_exploring_2009} and the intermediate Palomar Transient Factory \citep[iPTF;][]{kulkarni_intermediate_2013} surveys. Data were routinely collected by the Palomar 48-inch survey telescope in the Mould $R$ band \citep{law_palomar_2009}. Follow-up observations were conducted mainly with the robotic 60-inch telescope \citep{cenko_automated_2006} using an SDSS $r$-band filter, with additional telescopes providing supplementary photometry and spectroscopy \citep[see][]{gal-yam_real-time_2011}. We chose SNe that show hydrogen lines in their spectra (Type II), but do not show narrow emission lines at late times \citep[Type IIn;][]{schlegel_new_1990,filippenko_optical_1997,kiewe_caltech_2012}. This was done primarily because the optical emission from interacting SNe~IIn is dominated by their surrounding medium, and we are interested in the physics of the exploding star itself. We rejected transitional Type IIb SNe that develop strong He~I lines and resemble SNe~Ib. We also selected only SNe that had (1) at least five detections within ten days of the first detection, (2) well-sampled peaks/plateaus, and (3) an estimated date of explosion determined to within 3 days. 

The full list of SNe, their coordinates, and classification spectra is presented in Table \ref{tab:SNList}. Most of the spectra were obtained with the Double Spectrograph \citep{oke_efficient_1982} on the 5-m Hale telescope at Palomar Observatory, the Kast spectrograph \citep{miller_lick_1993} on the Shane 3-m telescope at Lick Observatory, the Low Resolution Imaging Spectrometer \citep[LRIS;][]{oke_keck_1995} on the Keck-1 10-m telescope, and the DEep Imaging Multi-Object Spectrograph \citep[DEIMOS;][]{faber_deimos_2003} on the Keck-2 10-m telescope. Spectral reductions followed standard techniques \citep[e.g.,][]{matheson_detailed_2000,silverman_berkeley_2012}. All spectra are publicly available via the Weizmann Interactive Supernova Data Repository \citep[WISeREP,][]{yaron_wiserep_2012}. 

\begin{figure}
	\centering
	\includegraphics[width=1\columnwidth]{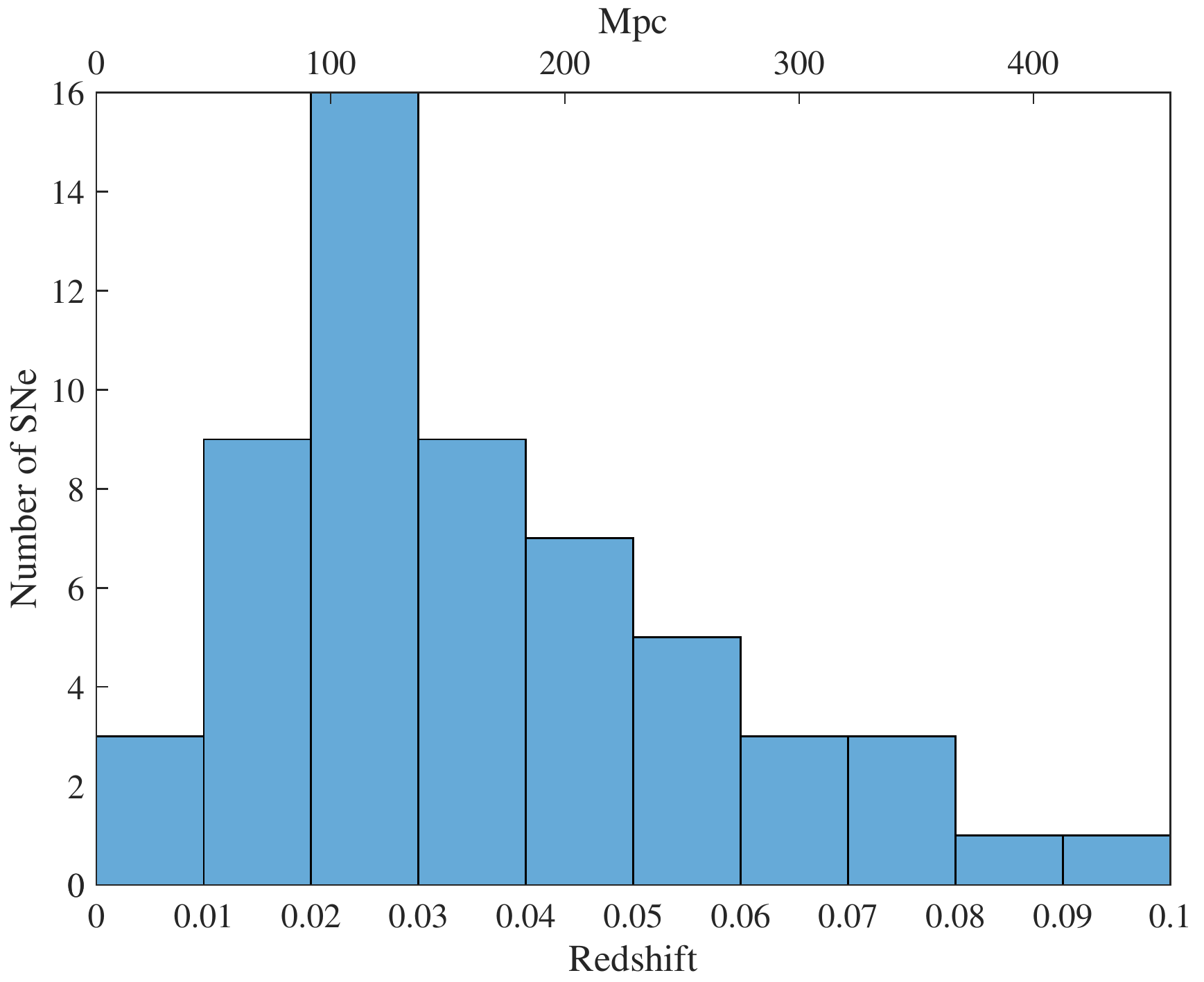}
	\caption{Redshift distribution of the SNe in the sample.}
	\label{fig:redshiftHist}
\end{figure}

The redshift ($z$) range is 0.0026--0.093, with a median value of 0.03. The distribution of redshifts is given in Figure \ref{fig:redshiftHist}. Note that this is a flux-limited survey, and is unbiased with respect to host galaxy. Some of the events in our sample briefly showed narrow emission lines which vanished after a few days. These are interpreted as ``flash-ionization events''(\citealt{gal-yam_wolf-rayet-like_2014}; Khazov et al. 2015, submitted). All of the photometry is available in the online material.

\begin{figure}
	\centering
	\includegraphics[width=1\columnwidth]{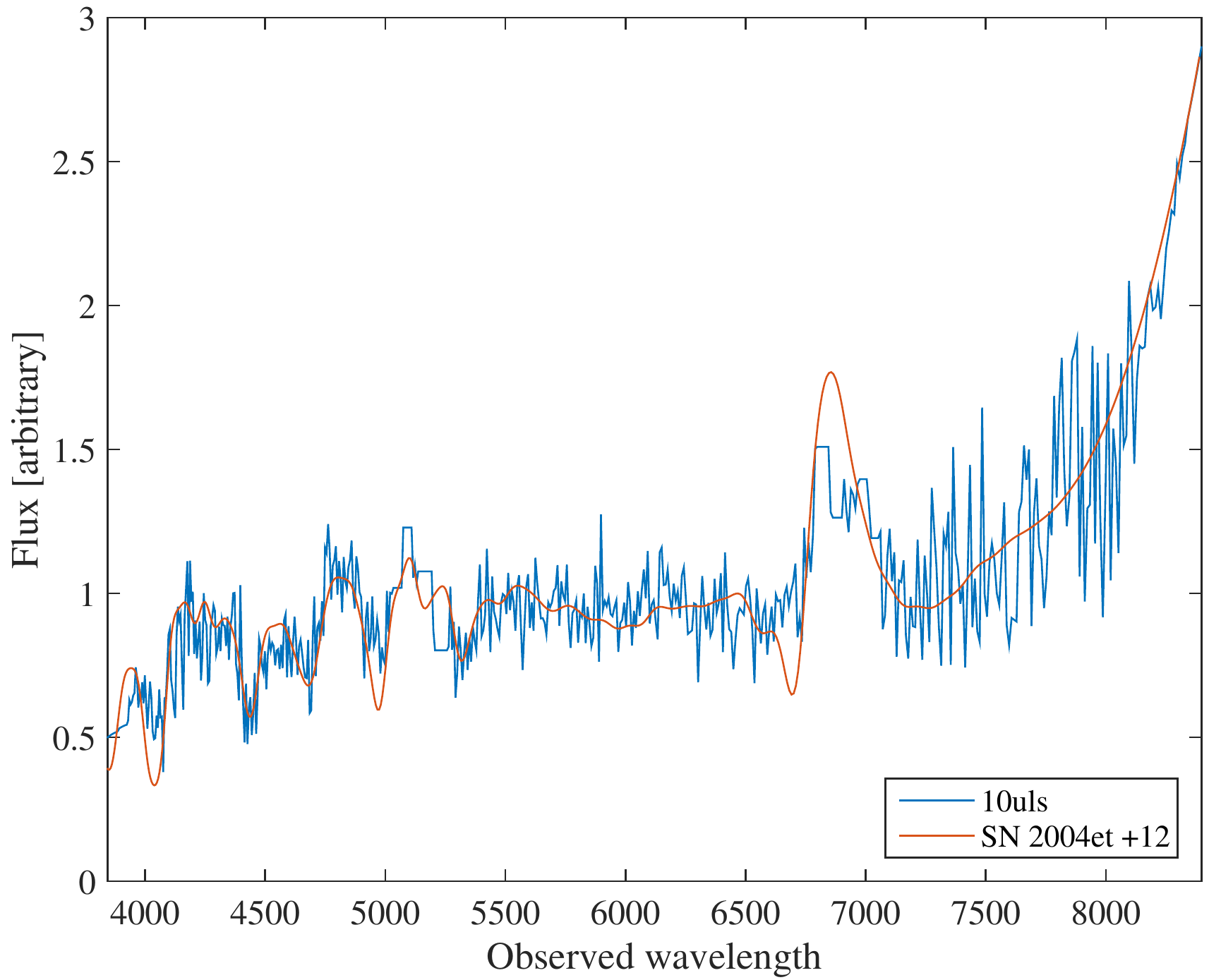}
	\caption{Spectrum of PTF10uls. Superimposed is the spectrum of SN 2004et (12 days after peak).}
	\label{fig:10ulsSpec}
\end{figure}

\begin{figure}
	\centering
	\includegraphics[width=1\columnwidth]{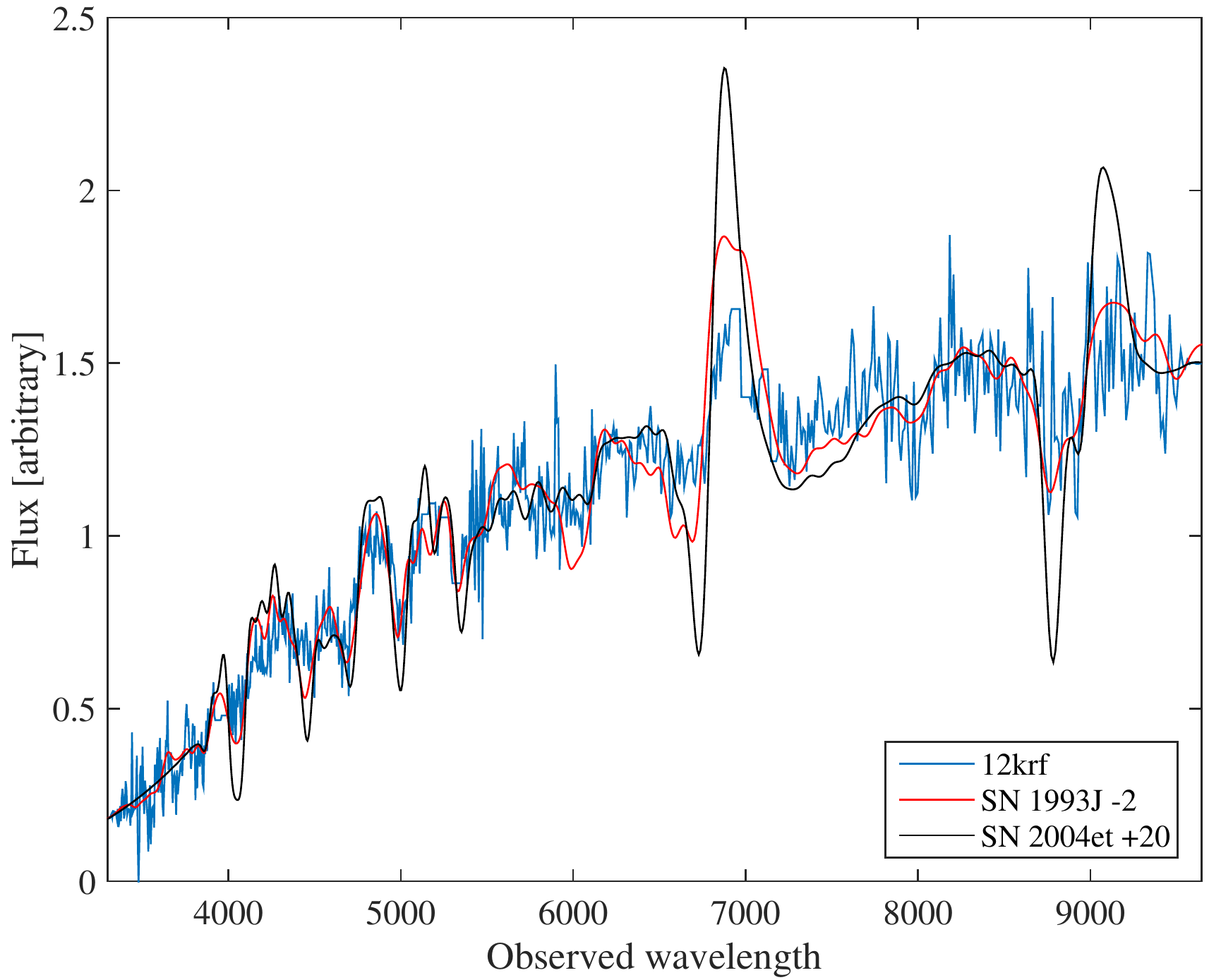}
	\caption{Spectrum of PTF12krf. Superimposed are the spectra of Type IIb SN 1993J (2 days before peak) and Type II-P SN 2004et (20 days after peak). Note the weaker He~I line compared to SN~IIb 1993J.}
	\label{fig:12krfSpec}
\end{figure}

\begin{figure}
	\centering
	\includegraphics[width=1\columnwidth]{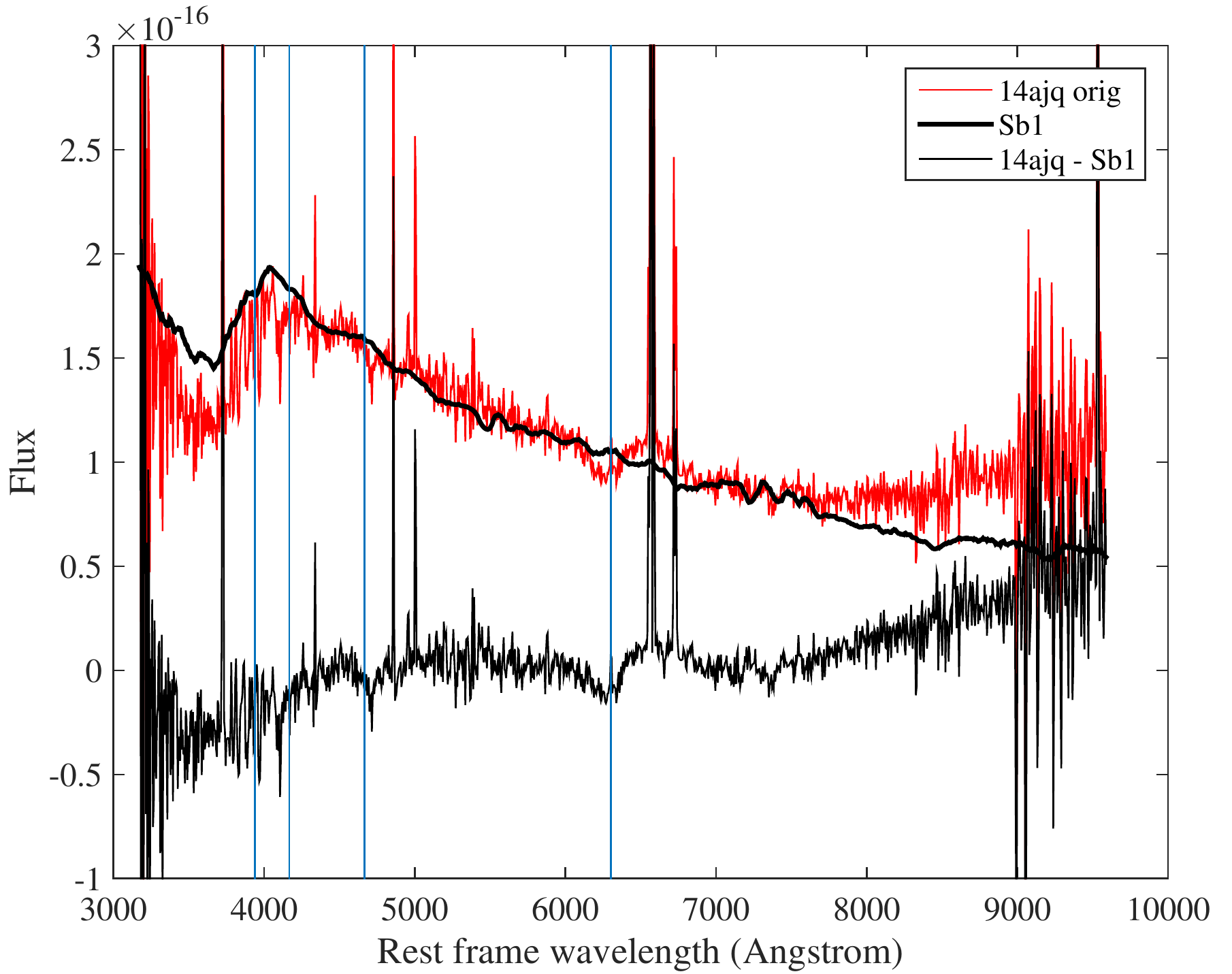}
	\caption{Subtraction of an Sb1 template from the spectrum of iPTF14ajq. The hydrogen Balmer series is shown offset by 12,000 km s$^{-1}$.}
	\label{fig:14ajqSpec}
\end{figure}

\cite{arcavi_continuum_2014} identified PTF10iam and PTF10nuj as abnormal transients. They were therefore discarded from the sample, leaving \numOfLCs{} events. All remaining objects had typical SN~II spectra. Three objects in the sample were difficult to classify but were ultimately retained. We compared the spectrum of PTF10uls and PTF12krf to templates using SNID \citep{blondin_determining_2007}. We found that PTF10uls is consistent with a SN~II-P spectrum (Figure \ref{fig:10ulsSpec}), while PTF12krf is consistent with a SN~II-P, though we cannot rule out that it is a SN~IIb (Figure \ref{fig:12krfSpec}). The spectrum of iPTF14ajq had significant galaxy contamination. Figure \ref{fig:14ajqSpec} shows the spectrum after subtraction of an Sb1 template \citep{kinney_template_1996}; it is that of a reddened SN~II.

\vfill\break
\section{Analysis}
\label{sec:analysis}

\subsection{Photometry}

The photometry was extracted using a point-spread function (PSF) fitting routine \citep{sullivan_photometric_2006,firth_rising_2015} applied after image subtraction. Photometry for iPTF13dkk, iPTF13dqy (Yaron et al. 2015; submitted to Nature Physics), and iPTF13dzb was supplemented with data from the Las Cumbres Observatory Global Telescope Network \citep[LCOGT;][]{brown_cumbres_2013}. This was obtained by PSF fitting, and fitting a low-order polynomial to the background. Photometry for PTF12cod was supplemented with data from the 40-inch telescope at Mount Laguna Observatory (MLO), which was also obtained with PSF fitting; see \cite{smith_ptf11iqb:_2015} for details on the MLO reduction procedure. Photometry for PTF10vdl (SN 2010id) was supplemented with that published by \cite{gal-yam_real-time_2011}, photometry for PTF12bvh (SN 2012aw) was supplemented with that published by \cite{munari_bvri_2013}, and photometry for PTF13aaz (SN 2013am) was supplemented with that published by \cite{zhang_optical_2014}. 

The light curves are presented in Appendix Figures \ref{fig:lcPlots1}---\ref{fig:lcPlots4}. We found that small additive constants (indicated in the figures) are needed to bring the supplementary and the 60-inch data in line with the PTF 48-inch observations; this is due most likely to the different $r/R$ filter responses of the 48-inch, 60-inch, and other data sources. The photometry was corrected for Galactic extinction using the \cite{schlegel_maps_1998} maps.\footnote{Derived using the \emph{sky\textunderscore ebv} routine in MATLAB, with $R_V=3.08$.} The distance moduli were calculated from the spectroscopic redshifts of the host galaxies using a cosmological model with H$_0=70$ km s$^{-1}$ Mpc$^{-1}$, $\Omega_m=0.3$, and $\Omega_\Lambda=0.7$. The analysis presented made use of the MATLAB$^{\circledR}$ package for astronomy and astrophysics \citep{ofek_matlab_2014}. 

The sample was not corrected for local host-galaxy extinction. \cite{faran_photometric_2014} explored various dust-extinction correction techniques including photometric methods based on comparison to low-extinction SNe, as well as spectroscopic methods using the Na~D doublet equivalent width. They found that none of the procedures increased the uniformity of their sample, and in some cases even increased the scatter. Thus, we would be introducing more uncertainty by correcting according to the classical prescriptions. However, we inspected the sample and found only five questionable objects. PTF09cjq has a red continuum, Na~D absorption lines, and likely suffers from extinction. PTF10bgl and iPTF14ajq have a red continuum, but no Na~D absorption lines, and may suffer from extinction. PTF10uqn, iPTF13bld, and iPTF13akg have a blue continuum, but show clear Na~D absorption lines, and may possibly suffer from extinction --- but this is unlikely to be significant, and it may be caused by host contamination in the spectra.

The time of explosion for most objects was estimated as the midpoint between the last non-detection and the first detection. For PTF09ecm, PTF10bgl, PTF10umz, PTF11iqb, PTF12efk, PTF12hsx, iPTF13cly, iPTF14adz, and iPTF14aoi --- where the limits were poorer, but the rise was well sampled --- we estimated the time of explosion with an exponential fit described in Section \ref{sec:lightCurveParamEstimation}. The fits are shown in Appendix Figure \ref{fig:expFits}.

The observed light curves were smoothed with a linear regression using a Gaussian kernel described in Section \ref{sec:lightCurveParamEstimation}. The full set of light-curve fits is shown in Appendix Figures \ref{fig:smoothedLC1}---\ref{fig:smoothedLC4}. We determined the time of maximum luminosity by a method similar to that used by \cite{gall_comparative_2015}. We fit a first-order polynomial to a three-day window of our smoothed light curve, and then shifted the window along the light curve. The algorithm was terminated when the slope of the polynomial surpassed $-0.01$ mag day$^{-1}$. The termination position of the algorithm was determined as the time of maximum luminosity. The change in magnitude between peak and 15 days post-peak, $\Delta m_{15}$ \citep[]{phillips_absolute_1993}, was determined by interpolating the smoothed light curve to 15 days after the time of maximum and subtracting the peak magnitude. These values are listed in Table \ref{tab:SNDerivedQuantities}.

\subsection{Comparison to RW11}
Care should be taken when comparing observations in the optical bands with RW11 or other models such as those of \cite{nakar_early_2010}, which converge at $t \approx 1$ day (note that all times given in this paper, unless stated otherwise, are relative to the estimated date of explosion and are given in the rest frame). The appropriate model has two strong requirements for validity: first, the emitting region must have originated in layers $\delta_m$ that were initially close to the surface of the star, $\delta_m \equiv (R_*-r)/R_* \ll 1$; second, the temperature must be greater than 1 eV, where Thomson scattering is dominant and recombination is negligible (see the beginning of Section 3 in RW11). Breakdown of the first assumption causes the dominant divergence of the solution by changing the final velocity of each element ($f_\nu$) from the asymptotic value used, $f_\nu=2$. This induces an \emph{underestimation} of the temperature (discussed in Section 3.1 of RW11), causing an overestimation of the luminosity, as can be seen in the model overshoot at later times of most of the fits presented in Appendix Figures \ref{fig:RWFits1}---\ref{fig:RWFits3}. 

Extending the fit to $t>4$ days forces the naturally overshooting region to coincide with the peak data. This has two effects: first, the $t<4$ day data get undershot; second, this procedure effectively shortens the rise time in the model, which reduces the probability of large radii. An example of the RW11 fit, and the effect of extending the fit beyond four days, is shown in Figure \ref{fig:RW11Examples}. 

As in \cite{gonzalez-gaitan_rise-time_2015} and \cite{gall_comparative_2015}, RSG models were compared to the sample without distinguishing between Type II-P and II-L SNe. This is justified for the following two reasons: first, there is evidence that Type II-P and II-L SNe may form a continuum in both their photometric and spectroscopic properties \citep[see][this work]{anderson_characterizing_2014,gutierrez_h_2014}, making it unlikely that they originate from different progenitors. Second, blue supergiants are unlikely progenitors primarily because their small radii \citep{kudritzki_quantitative_2008} will cause severe adiabatic losses, and they will not have the energy budget to reach the peak luminosity of Type II-L SNe which can peak above -18 (Figure \ref{fig:empiricalCorrs}). Yellow hypergiants are extremely rare \citep{oudmaijer_post-red_2009}, and they are known to be associated with Type IIb SNe \citep{maund_yellow_2011}. This leaves RSGs as a reasonable default for the progenitors of Type II-P and II-L SNe.

In order to compare with RW11 models, we selected only those events with at least five detections in the first four days from explosion; this left 33 events. We then generated bolometric light curves using Equation 14 of RW11 (appropriate for RSGs, $n=3/2$):
\begin{equation}
	L = 8.5\times 10^{42} \frac{E^{0.92}_{51} R_{*,13} }
	{f^{0.27}_{\rho} (M/{\rm M}_{\odot})^{0.84}\kappa^{0.92}_{0.34} }t^{-0.16}_5 \; \text{erg} \, {\rm s}^{-1},
	\label{eq:RW11-L}
\end{equation}

\n where the explosion energy $E_{51} = E/10^{51}$ erg, the progenitor radius $R_{*,13} = R/10^{13}$ cm, the opacity $\kappa_{0.34} = \kappa / 0.34$ g$^{-1}$ cm$^2$, the ejecta mass $M$, and the time from explosion $t_5 = t / 10^5$ s. Also, $f_\rho \equiv \rho_{1/2} / \bar{\rho}$, where $\bar{\rho}$ is the mean density of the ejecta and $\rho_{1/2}$ is the density at $r=R_*/2$. Note that $n$ is the index of the density at the edge of the ejecta given by $\rho(r_0) = \rho_{1/2} \delta^n$. The apparent $R$-band magnitude was calculated with the photospheric temperature given in Equation 13 of RW11,

\begin{equation}
	T_{\rm ph} = 1.6 f_\rho^{-0.037} \frac{E_{51}^{0.027} R_{*,13}^{1/4}}{(M/M_\odot)^{0.054}\kappa_{0.34}^{0.28}}t_5^{-0.45} \; \text{eV,}
\end{equation}

\n corrected to the color temperature with the factor $T_{\rm c} = 1.2\, T_{\rm ph}$ (see discussion in RW11, their Section 3.2 and Figure 1). Then the modeled $R$-band magnitudes were calculated with the \emph{synphot} routine \citep{ofek_matlab_2014}.

We generated light curves for a grid of RSG progenitors with radii $R_* = 10^2$--$10^3$ \; R$_\odot$ (200 points logarithmically spaced), explosion energies $E_{51} \equiv E / 10^{51} \; \text{erg} = 10^{-2}$--$10^2$ (250 points logarithmically spaced), a fixed ejected mass $M_{10} \equiv M /(10 \; M_\odot) = 1$,\footnote{The early-time light curve depends on the energy per unit mass; therefore, the possible diversity in $M$ is covered by our range of $E_{51}$. For further discussion see \cite{rabinak_early_2011} and \cite{ganot_detection_2014}.} $f_\rho=0.1$, $\kappa_{0.34} = 1$, and various times of explosion within our uncertainty on this date (50 points linearly spaced between $t_0\pm\Delta t_0$).

With the model light curves in hand, we calculated the $\chi^2$ values for the observed flux (in the first 4 days from explosion) for all combinations of the radius, explosion energy per unit mass, and all possible dates of explosion. Finally, we scaled the flux errors until the minimal $\chi^2$ reached 1.\footnote{The flux errors from our pipeline are underestimated, leading to high $\chi^2$ for models which fit the data well. The errors were scaled to allow for the comparison of different models to each other. The scaling values are presented in Appendix Figures \ref{fig:RWFits1}---\ref{fig:RWFits3}.} The energy per unit mass was estimated at the minimum $\chi^2$ of the grid, and the 95\% confidence interval was estimated using a profile likelihood: finding optimal $t_0$ and $R_*$ for each energy, and then finding values of $E_{51}$ where the cumulative distribution function CDF$(\chi^2)\leq 0.95$. The $E/M$ values we determined are listed in Table \ref{tab:SNDerivedQuantities}, and are shown in Figure \ref{fig:RW11Energies}.

To calculate the mean value of $E/M$, we have to correct for Malmquist bias. We used an effective distance modulus $DM^*$ such that all SNe would have the same peak apparent magnitude as the faintest SN in our sample with the formula

\begin{equation}
	\text{Max}\{m^{\text{peak}}\} = M^{\text{peak}}_i + DM^*_i,
\end{equation}

\n where Max$\{ m^{\text{peak}} \}$ is the faintest peak apparent magnitude in the sample, $DM^*_i$ is the effective distance modulus that sets the peak absolute magnitude $M^{\text{peak}}_i$ equal to the faintest peak apparent magnitude. Then the mean SN $E/M$ value in the sample was calculated as

\begin{equation}
	\left\langle \frac{E}{M} \right\rangle = \frac{\Sigma_i \left( E/M \right)_i / D^{*3}_i}{\Sigma_j 1/D^{*3}_j},
\end{equation}

\n where $D^*$ is the luminosity distance taken from the relation

\begin{equation}
	DM^* = 5 \log_{10} \left( \frac{D^*}{10 \; \text{pc}}\right).
\end{equation}

\n This procedure accounts for the overrepresentation of luminous events in our flux-limited sample. By weighting according to their equivalent volume, less-luminous events --- which naturally have a small volume --- get put on equal footing with more-luminous events. The corrected histogram of $E/M$ values is shown in Figure \ref{fig:RW11Energies}.

\subsection{Spectroscopy}

We estimate the expansion velocity of each SN by measuring the minimum of the H$\alpha$ P-Cygni profile. This was accomplished by fitting a second-order polynomial to the H$\alpha$ absorption. In order to normalize the velocities to a uniform epoch, the relation from \cite{faran_photometric_2014} was used, relating the velocities measured to the velocity on day 50 for SNe~II-P. The relation is given by

\begin{equation}
	v_{50} = v_{{\rm H}{\alpha}}(t)\left(\frac{t}{50}\right)^{0.412 \pm 0.02}.
\end{equation}

\n The measured velocities are presented in Table \ref{tab:SNDerivedQuantities}.

\subsection{\Ni{} Mass Estimation}

\n For eight events which had good late-time coverage, we fit for the synthesized radioactive nickel mass. The luminosity per unit mass released by radioactive $^{56}$Ni is given by

\begin{eqnarray}
	l & = & 3.9\times 10^{10} e^{ -t / \tau_{\rm Ni} } + \\
	& & 7 \times 10^9 \left( e^{ -t / \tau_{\rm Co} } - e^{ -t / \tau_{\rm Ni} } \right) \; \text{erg} \; \text{g}^{-1} \; \text{s}^{-1} , \nonumber
\end{eqnarray}

\n where $\tau_{\rm Ni}$ and $\tau_{\rm Co}$ are 8.8, and 111.09 days, respectively. For each of the relevant SNe, we fit the initial nickel mass by minimizing the linear least-squares equation

\begin{equation}
	L(t_i) = M_{\rm Ni} l(t_i),
\end{equation}

\n where $L(t_i)$ and $l(t_i)$ are (respectively) the observed luminosity and expected luminosity per unit mass at time $t_i$, and $M_{\rm Ni}$ is the initial nickel mass. We also fit three events from the literature for which the authors derived \Ni{} masses using multi-band quasi-bolometric light curves (SN 2005cs, SN 2012aw, and SN 2013ab; Figure \ref{fig:Ni56VE51}), and found that our values for $M_{\rm Ni}$ are sufficiently close to justify no bolometric correction. However, to be conservative, we assume a 50\% uncertainty in our derived \Ni{} mass. These values are reported in Table \ref{tab:SNDerivedQuantities}.

\section{Results}

\label{sec:results}

\begin{figure*}[ht]
	\centering
	\includegraphics[width=1\textwidth]{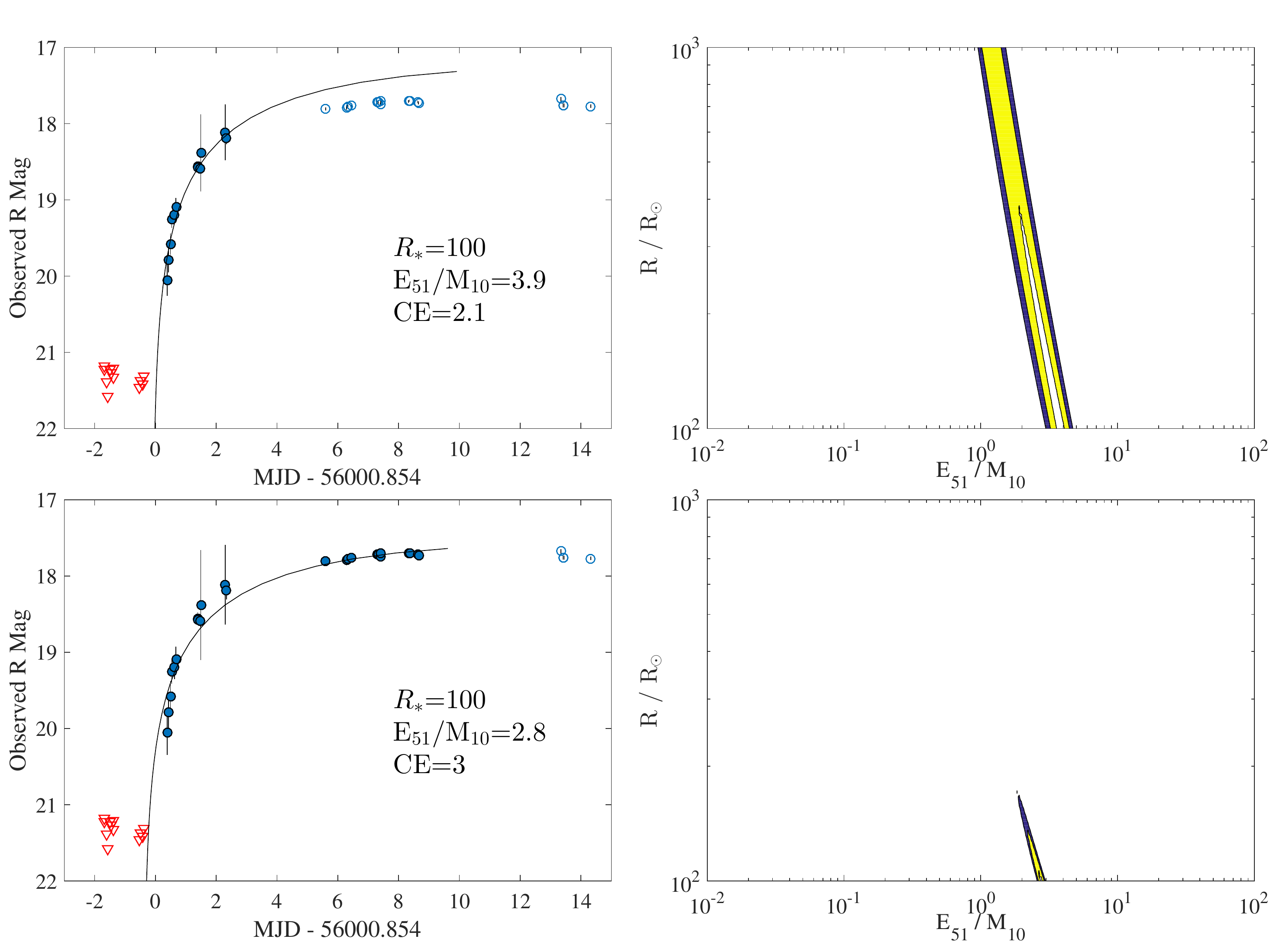}
	\caption{Top left: Example RW11 best fit to PTF12bro. The values of the progenitor radius $R$, explosion energy per unit mass $E_{51}/M_{10}$, and error scaling factor CE are displayed. Only filled symbols were included in the fit. Top right: The projection of $\chi^2$ onto the $R-E$ plane (at optimal explosion date $t_0$ for each point). The contours of 68\%, 95\%, and 99.7\% confidence intervals are shown. Bottom left, right: Best fit to PTF12bro and $\chi^2$ contours including data where $t \leq 10$. Notice that including later data reduces the probability for higher radii. \label{fig:RW11Examples}}
\end{figure*}

Figure \ref{fig:RW11Examples} shows an example of the fit to a RW11 model. We found that RW11 models describe the early-time light curves well in most cases (Appendix Figures \ref{fig:RWFits1}---\ref{fig:RWFits3}). For each combination of $E/M$ and $R_*$ the time of explosion was selected that minimizes the $\chi^2$. The contours represent the 68\%, 95\%, and 99.7\% $\chi^2$ confidence intervals. Figure \ref{fig:RW11Examples} is typical, and demonstrates that while the radius of the progenitor cannot be constrained based on the early-time $R$-band light curve, $E/M$ can be estimated to better than a factor of five.

\begin{figure*}[ht]
	\centering
	\includegraphics[width=1\textwidth]{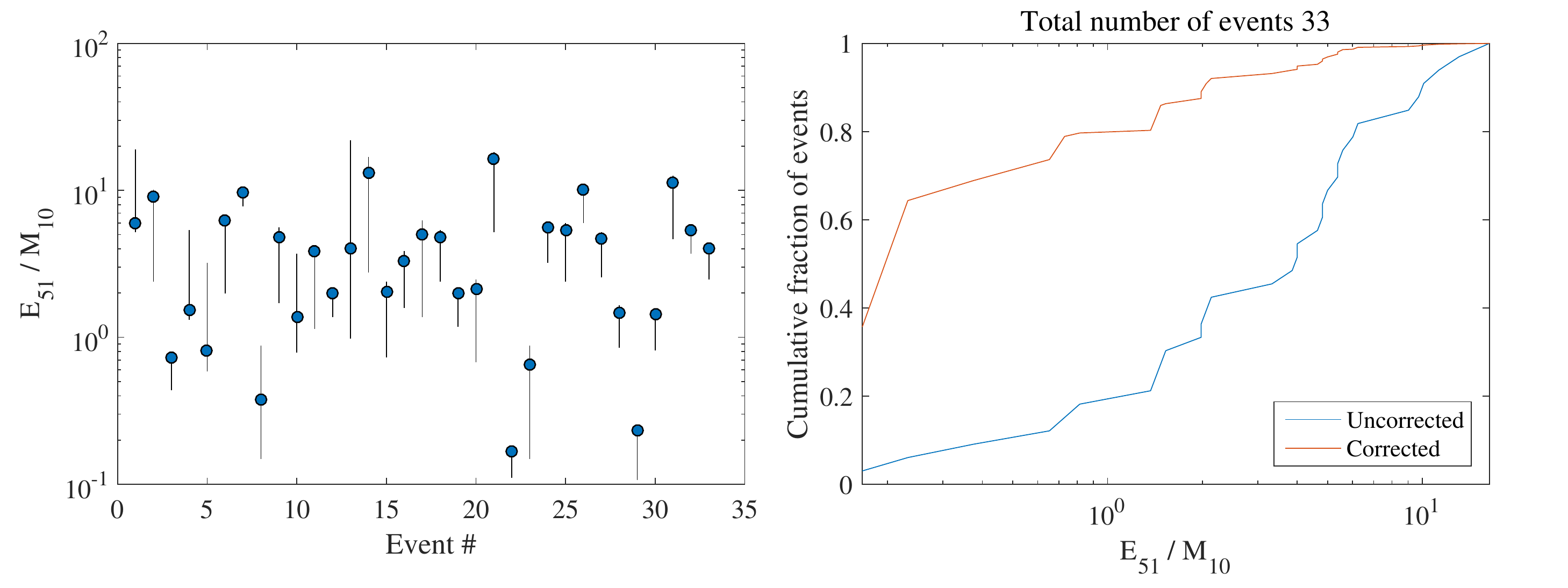}
	\caption{Left: $E/M$ derived from RW11 with 95\% profile likelihood confidence intervals. Right: Cumulative fraction of events below a given $E/M$, corrected and uncorrected for Malmquist bias. \label{fig:RW11Energies} }

\end{figure*}

\begin{figure*}[ht]
	\centering
	\includegraphics[width=1\textwidth]{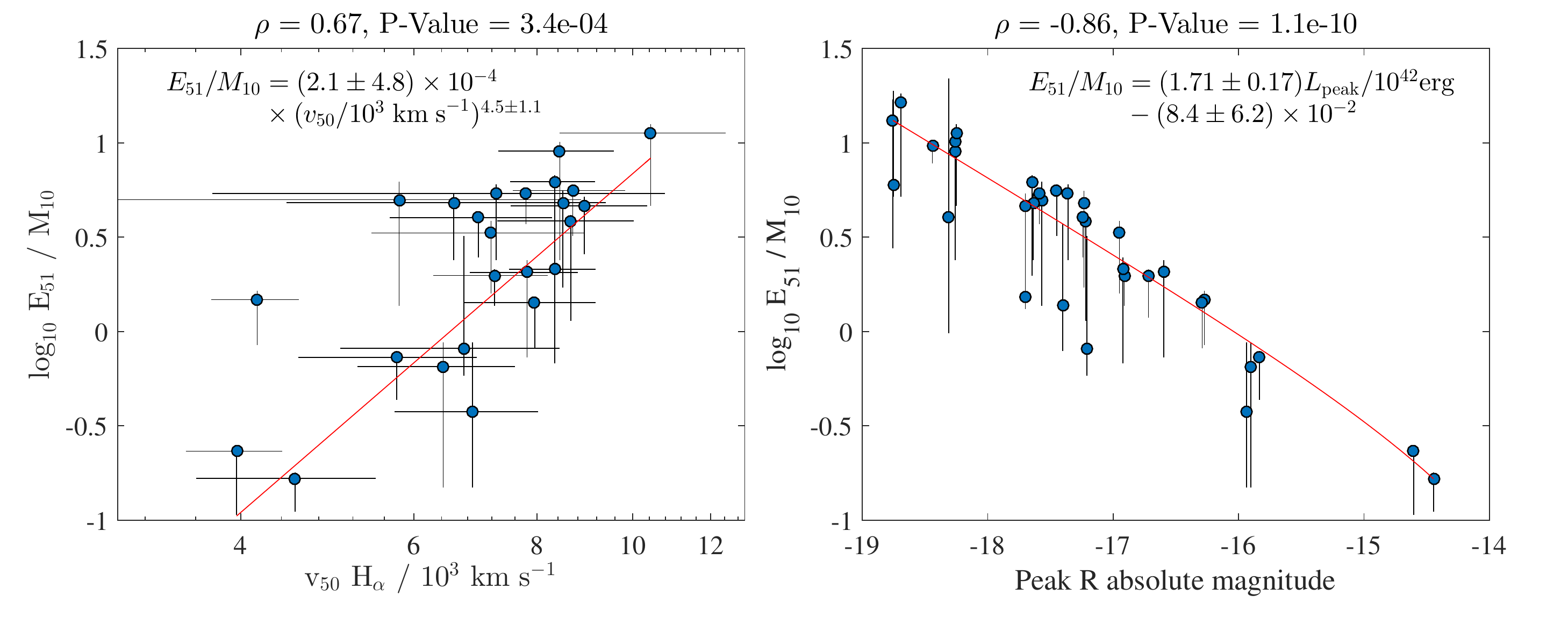}
	\caption{Left: $E/M$ from the fit to RW11 as a function of the velocity normalized to $v_{50}$ of H$\alpha$.
	Right: $E/M$ from RW11 as a function of peak magnitude. Red lines are the best fits described in the text. \label{fig:e51Corr}}
\end{figure*}

The energies derived for each SN and the cumulative fraction of events below a given $E/M$ are shown in Figure \ref{fig:RW11Energies}. We find that $E/M$ spans a range of $\sim (0.2-20)\times 10^{51}$ erg/(10 M$_\odot$). Moreover, the $E/M$ values deduced from the fit to RW11 models are significantly (P-value $<< 0.05$) correlated with the observed photospheric velocity (Figure \ref{fig:e51Corr}).\footnote{All correlations reported were calculated using the Spearman correlation test.} Taking the confidence interval as symmetric, a power-law fit gives

\begin{eqnarray}
	E_{51}/M_{10} &=&(2.1\pm4.8)\times 10 ^{-4} \\
	& &\times (v_{50}/10^3~ \text{km s}^{-1})^{4.5 \pm 1.1}, \nonumber
\end{eqnarray}

\n where the uncertainties are 95\% confidence intervals. We find that $E/M$ from the fit is also significantly correlated with the peak magnitude (Figure \ref{fig:e51Corr}), and is related to the peak luminosity by

\begin{eqnarray}
	E_{51}/M_{10} &=& (1.71 \pm 0.17) L_\text{peak}/10^{42}~ \text{erg}\\
	& & -(8.4 \pm 6.2)\times 10^{-2} \nonumber,
\end{eqnarray}

\n where $L_\text{peak}$ is the peak luminosity.

\begin{figure}[ht]
	\centering
	\includegraphics[width=1\columnwidth]{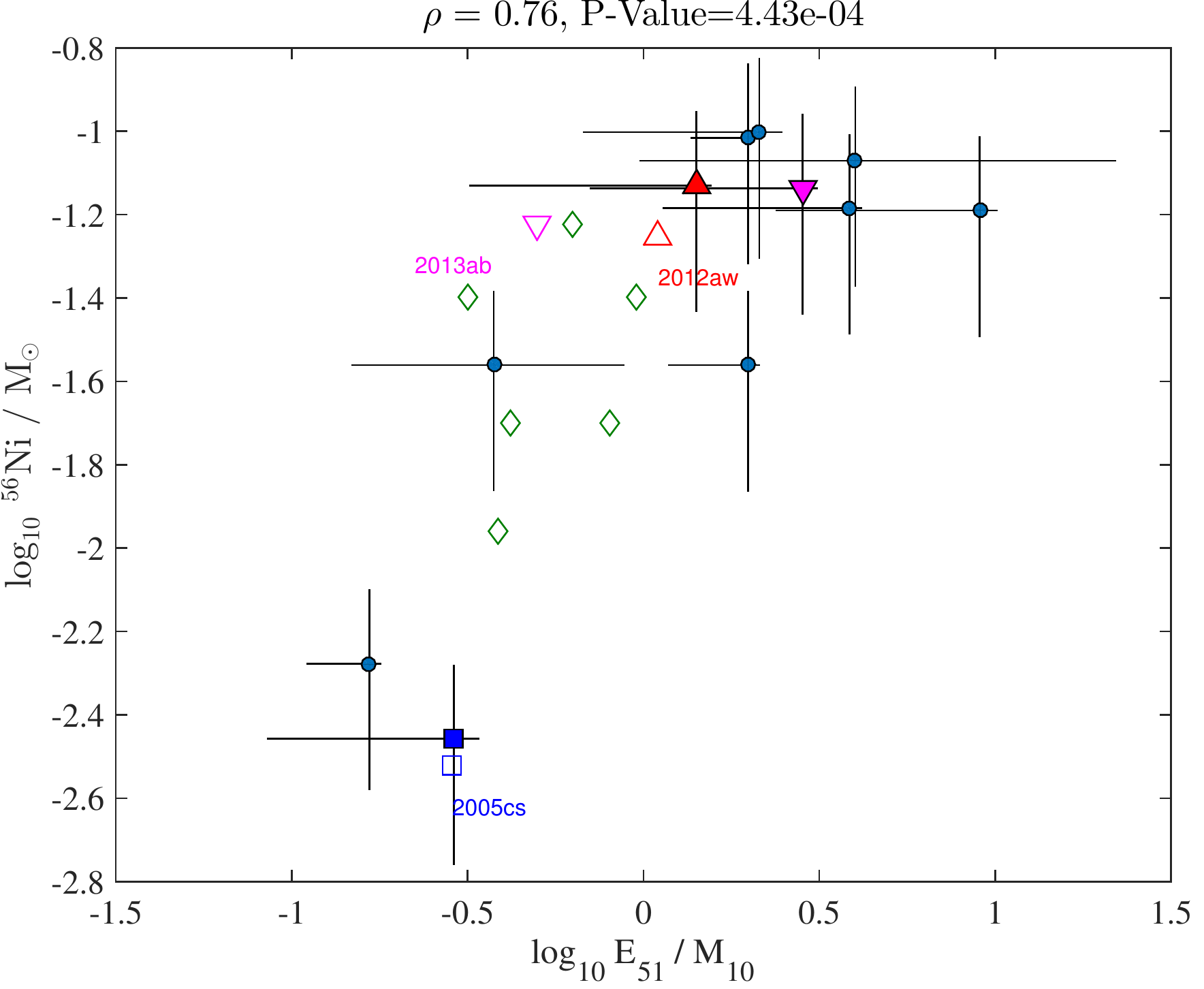}
	\caption{$E/M$ from the fit to RW11 vs. $^{56}$Ni mass. Empty symbols are taken from the literature (Table \ref{tab:Ni56MassLit}) with parameters estimated from hydrodynamic modeling. The blue, red, and magenta filled (empty) symbols represent our (literature) estimates of the parameters of SN 2005cs, SN 2012aw, and SN 2013ab. Note the good agreement in \Ni{} mass between our analysis and the literature. The source of the discrepancy between our $E/M$ estimate for SN 2013ab and that of \cite{bose_sn_2015} is unclear, due to the different methods used. Diamonds represent events from the literature which did not have sufficient early-time data on which to perform our analysis.}
	\label{fig:Ni56VE51}
\end{figure}

We added to our sample several events from the literature which have determined parameters from hydrodynamic light-curve modeling (see Table \ref{tab:Ni56MassLit}). Three SNe (SN 2005cs, SN 2012aw, and SN 2013ab) were sufficiently well sampled during their rise to allow us to perform our RW11 analysis, although it was necessary to slightly relax our criteria and include $R$-band data up to day 6 from explosion. We found our results to be consistent with the estimated explosion parameters from the literature (Figure \ref{fig:Ni56VE51}); however, we derive a higher $E/M$ value for SN 2013ab than do \cite{bose_sn_2015}. \cite{bose_sn_2015} estimated $E/M$ from hydrodynamic modeling, making it difficult to assess the source of this discrepancy. Note that our derived \Ni{} mass of $\sim 5\times 10^{-3}$ M$_\odot$ for iPTF13aaz (SN 2013am) is lower than the $1.5 \times 10^{-2}$ M$_{\odot}$ reported by \cite{zhang_optical_2014}, but the source of this discrepancy is unclear. We find that the \Ni{} mass is strongly correlated with $E/M$ ($\rho=\NiPValue{}$, P-value $<<0.05$; Figure \ref{fig:Ni56VE51}). This result has been observed in Type Ib/c SNe \citep{mazzali_very_2013}, and is in line with models such as that of \cite{kushnir_thermonuclear_2015}, which predict that more \Ni{} is produced by more-energetic SN~II explosions.

\begin{figure*}[ht]
	\centering
	\includegraphics[width=1\textwidth]{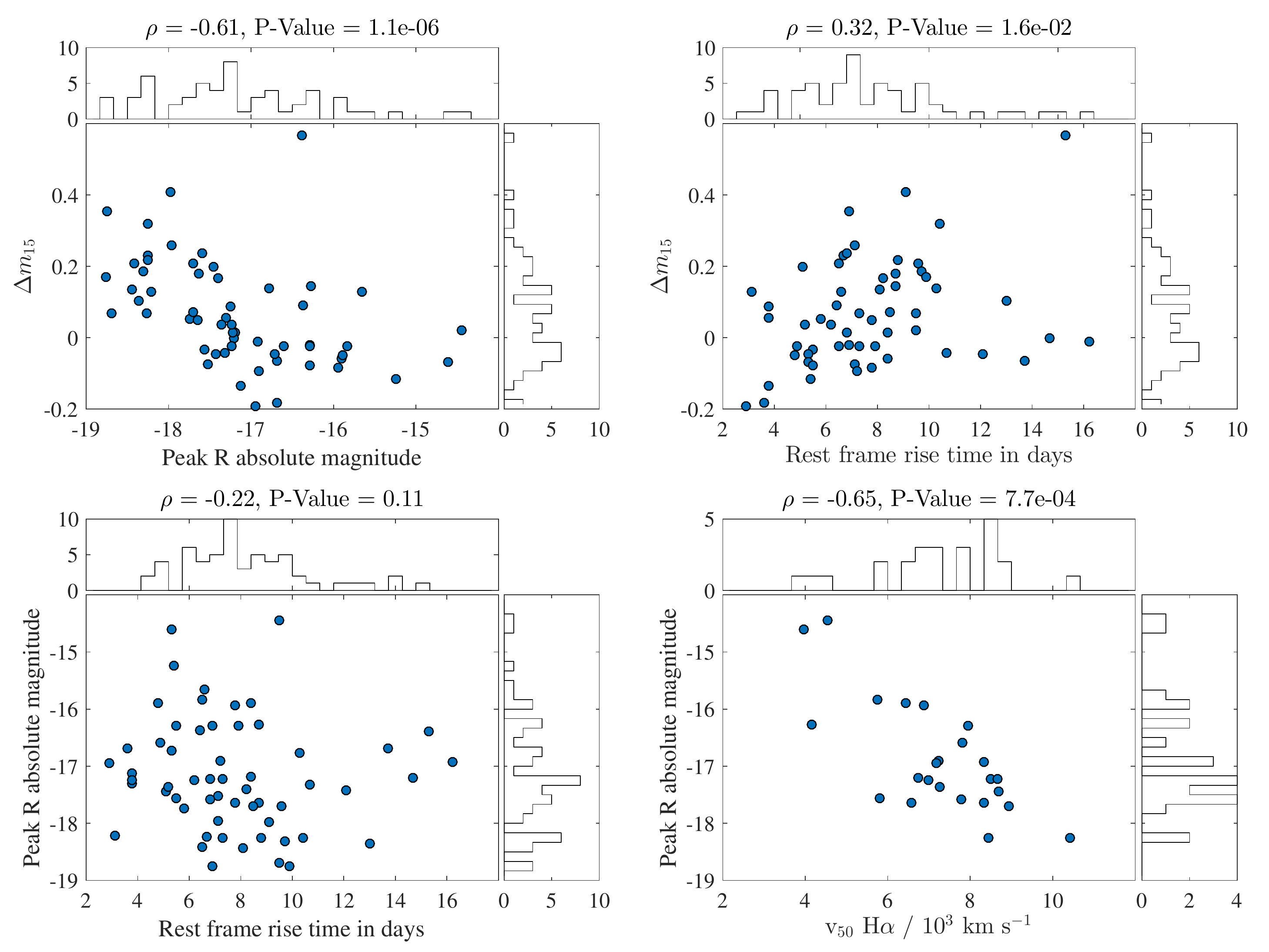}
	\caption{Top left: $\Delta m_{15}$ as a function of the peak magnitude. Top right: $\Delta m_{15}$ as a function of rise time. Bottom left: Peak magnitude as a function of rise time, derived from the smoothed light curves (Appendix Figures \ref{fig:smoothedLC1}---\ref{fig:smoothedLC4}). Bottom right: Peak magnitude as a function of $v_{50}$. \label{fig:empiricalCorrs}}
\end{figure*}

In addition, we find several empirical correlations (Figure \ref{fig:empiricalCorrs}). The peak luminosity is significantly and strongly correlated with $\Delta m_{15}$ --- brighter events decline faster. This is the opposite of well-established trends in SNe~Ia and Ib/c \citep{phillips_absolute_1993} that are powered by $^{56}$Ni during their rise, and is in agreement with the findings of \cite{anderson_characterizing_2014} for Type II $V$-band light curves. The peak luminosity is also correlated with $v_{50}$: brighter events have higher velocities at day 50. This relation has already been established for Type II-P SNe \citep{hamuy_type_2002,nugent_toward_2006}, although it has not been demonstrated until now for SNe~II generally. The rise time is more weakly correlated with $\Delta m_{15}$, and with a larger scatter, although it significantly shows that slower risers are also faster decliners. We do not observe a significant correlation between the rise time and the peak luminosity, contrary to the suggestions of \cite{gal-yam_real-time_2011} and \cite{gall_comparative_2015}.

\section{Discussion}

\label{sec:discussion}

We have performed the first direct fitting of analytical early light-curve models (RW11) to a large sample of Type II SNe with a well-sampled rise. Our results show that, assuming a RSG progenitor, we can deduce the value of $E/M$ to within roughly a factor of five from early-time optical light curves. Progenitor radii are not constrained by $R$-band data alone, and require UV observations \citep{ganot_detection_2014}. 

\cite{gall_comparative_2015} and \cite{gonzalez-gaitan_rise-time_2015} recently compared light curves to RW11/NS10 shock-cooling models and found that only small radii ($R < 400$ R$_\odot$) appeared to be consistent with observations, in strong tension with direct measurements of Type II-P SNe \citep{smartt_death_2009,smartt_observational_2015}. However, \cite{valenti_first_2014} and \cite{bose_sn_2015} compared single objects and found no such discrepancy. 

It appears that the \cite{gall_comparative_2015} and \cite{gonzalez-gaitan_rise-time_2015} method of extracting a rise time from the models and comparing it to the rise time of their light curves is inaccurate. The models are valid for a brief ($t\approx 4$ days) period before important assumptions such as the emission coming from a thin shell at the edge of the star, and the temperature being well above 1 eV, begin to break down. As was explained in Section \ref{sec:analysis}, the breakdown of these assumptions leads to an overshoot of the data at later times ($t>4$ days). By comparing to rise times from the models, these works rejected models which fit the early photometric data well. 

\begin{figure*}
	\centering
	\includegraphics[width=1\columnwidth]{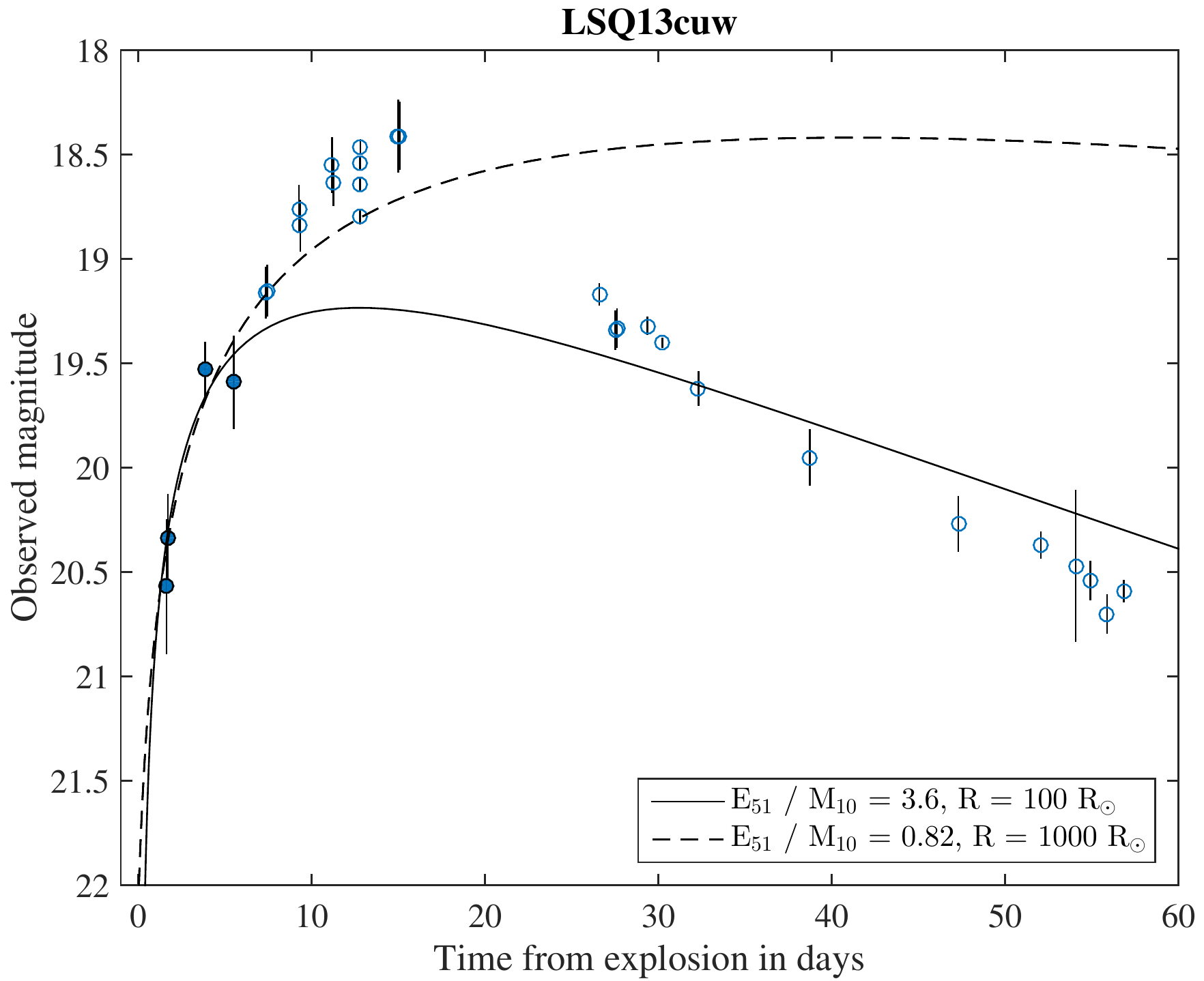}
	\caption{Comparison of RW11 models to LSQ13cuw photometry from \cite{gall_comparative_2015}. Note that while they differ greatly in rise time, models with progenitor radii of 100--1000 \; R$_\odot$ are consistent with the early-time measurements. Data up to day 6 were included in the fit (filled symbols).}
	\label{fig:comparisonToGall}
\end{figure*}

Figure \ref{fig:comparisonToGall} demonstrates this using LSQ13cuw, a well-sampled event from \cite{gall_comparative_2015}. We fit two extreme cases, with radii of 100 and 1000 R$_\odot$, to the first six days of data. Both models fit the early-time data equally well. While the 100 R$_\odot$ model has a consistent rise time with LSQ13cuw, it does not agree at all with the photometry near peak. We suspect that this explains the apparent discrepancy between the radii inferred from the models by \cite{gall_comparative_2015} and \cite{gonzalez-gaitan_rise-time_2015}, and the measured RSG radii \citep{smartt_death_2009} of the progenitors of SNe~II-P.

We find a strong correlation between the RW11 $E/M$ values and the SN expansion velocity at day 50. Because $v_{50}$ is an independent estimate of $E/M$, this provides support for the deduced $E/M$ values. We find that our sample has a mean energy per unit mass, corrected for Malmquist bias, of $\left\langle E_{51}/M_{10} \right\rangle = 0.85$, with a range of $E_{51}/M_{10} \approx 0.2$--20. Because the progenitor mass of a SN II-P is suggested to be confined to a relatively narrow range \citep[8--16 \; M$_\odot$;][]{smartt_death_2009}, our results lead to the conclusion that there is a significant intrinsic diversity in explosion energies. The correlation between peak magnitude and $E/M$ indicates that more energetic explosions also have higher peak luminosity. In addition, the strong positive correlation between $E/M$ and \Ni{} mass implies that stronger explosions produce more \Ni{}. This result is consistent with the predictions of some models, including those of \cite{kushnir_failure_2015} and \cite{kushnir_thermonuclear_2015}, claiming that the explosion mechanism of CC~SNe is thermonuclear detonation of the infalling outer shells.

In our sample, we do not find that the rise time and peak magnitude of SNe~II are correlated (Figure \ref{fig:empiricalCorrs}). Although it was suggested in the past that brighter SNe~II-P may have longer rise times \citep{gal-yam_real-time_2011}, our sample of well-monitored light curves disfavors this hypothesis. Our correlation between $\Delta m_{15}$ and the peak magnitude recovers a relation previously shown by \cite{anderson_characterizing_2014} in the $V$ band. We also find, however, that $\Delta m_{15}$ is correlated with the rise time, although with a large scatter. SNe with longer rise times also decline faster. 

\cite{nicholl_diversity_2015} have recently explored various mechanisms to explain hydrogen-poor superluminous SNe \citep[SLSNe,][]{gal-yam_luminous_2012}. Their models include magnetars \citep{kasen_supernova_2010,woosley_bright_2010}, circumstellar interaction \citep{woosley_pulsational_2007,ofek_supernova_2010,chevalier_shock_2011}, and \Ni{} radioactive decay. They find an opposing correlation to ours: in hydrogen-poor SLSNe, as well as in all of the above-mentioned models, longer-rising SNe also decline more slowly. We interpret this as evidence that Type II SNe are not powered by any of these potential sources during their early phase.

\section{Summary}

Our main conclusions regarding SNe~II can be summarized as follows.

\begin{itemize}

	\item The progenitor radius cannot be inferred by comparison to shock-cooling models based on $R$-band photometry alone. The value of $E/M$ can be inferred to within a factor of five.

	\item The mean SN~II energy per unit mass, corrected for Malmquist bias, is $\left \langle E/M \right \rangle = 0.85\times 10^{51}$ erg/(10 M$_\odot$), and has a range of (0.2--20) $\times 10^{51}$ erg/(10 M$_\odot$).

	\item The derived value of $E/M$ from RW11 models is strongly correlated with the photospheric velocity at day 50, peak magnitude, and \Ni{} mass produced in the explosion.

	\item $\Delta m_{15}$ is correlated with the rise time --- slower risers are also faster decliners. This indicates that Type II SNe are unlikely to be powered by radioactive decay or other central-engine models at early times.

\end{itemize}

While it was not possible to infer the radius from $R$-band data alone, the path for future work is clear. Multi-band light curves, which will be acquired by future surveys such as the Zwicky Transient Facility \citep[ZTF;][]{bellm_zwicky_2014,smith_zwicky_2014} and the Large Synoptic Survey Telescope \citep[LSST;][]{ivezic_lsst:_2008}, as well as early-time UV photometry from satellites such as ULTRASAT \citep{sagiv_science_2014}, will drastically reduce the uncertainties in determining the progenitor radius. The benefit will be twofold: these facilities will reduce uncertainties in the time of first light, and there will be more useful photometry within the window of validity of available shock-cooling models, because the rise time is much shorter in blue and UV bands.

\acknowledgments

We are grateful to the staffs at the many observatories where data for this study were collected (Palomar, Lick, Keck, etc.). We thank J. Silverman, G. Duggan, A. Miller, A. Waszczak, E. Bellm, K. Mooley, J. Van Roestel, A. Cucchiara, R. J. Foley, M. T. Kandrashoff, B. Sesar, I. Shivvers, J. S. Bloom, D. Xu, J. Surace, and L. Magill for helping with some of the observations and data reduction. Specifically for assistance with the MLO observations, we acknowledge N. Duong, T. Fetherolf, S. Brunker, R. Dixon, and A. Rachubo.

A.G.Y. is supported by the EU/FP7 via ERC grant no. 307260, the Quantum Universe I-CORE Program by the Israeli Committee for Planning and Budgeting and the Israel Science Foundation (ISF); by Minerva and ISF grants; by the Weizmann-UK ``making connections'' program; and by
Kimmel and ARCHES awards.
E.O.O. is incumbent of the Arye Dissentshik career development chair and is grateful for support by grants from the Willner Family Leadership Institute Ilan Gluzman (Secaucus, NJ), ISF, Minerva, Weizmann-UK, and the I-CORE Program of the Planning and Budgeting Committee and the ISF.
M.S. acknowledges support from the Royal Society and EU/FP7-ERC grant no [615929].
K.M. is grateful for a Marie Curie Intra-European Fellowship, within the 7th European Community Framework Programme (FP7).
D.C.L., S.F.A., J.C.H., and J.M.F. are supported by NSF grants AST-1009571 and AST-1210311, under which part of this research (photometry data collected at MLO) was carried out. 
The supernova research of A.V.F.'s group at U.C. Berkeley presented herein is supported by Gary \& Cynthia Bengier, the Christopher R. Redlich Fund, the TABASGO Foundation, and NSF grant AST-1211916.

Research at Lick Observatory is partially supported by a generous gift from Google. Some of the data presented herein were obtained at the W. M. Keck Observatory, which is operated as a scientific partnership among the California Institute of Technology, the University of California, and NASA; the observatory was made possible by the
generous financial support of the W. M. Keck Foundation. The William Herschel Telescope is operated on the island of La Palma by the Isaac Newton Group in the Spanish Observatorio del Roque de los Muchachos of the Instituto de Astrof\'isica de Canarias. This research has made use of the APASS database, located at the AAVSO web site. Funding for APASS has been provided by the Robert Martin Ayers Sciences Fund. A portion of this work was carried out at the Jet Propulsion Laboratory under a Research and Technology Development Grant, under contract with the National Aeronautics and Space Administration. Copyright 2015 California Institute of Technology. All Rights Reserved. US Government Support Acknowledged. LANL participation in iPTF is supported by the US Department of Energy as part of the Laboratory Directed Research and Development program.

\bibliographystyle{apj}
\bibliography{biblio.bib}

\begin{deluxetable*}{lllllllll}
	\tablewidth{0pt} 
	\tablecolumns{9}
	\tablecaption{List of supernovae included in this study. \label{tab:SNList}}
	\tablehead{  % column headings
		\colhead{}& \colhead{}& \colhead{}&\colhead{} &\colhead{} &\colhead{} &\colhead{} & \multicolumn{2}{l}{Classification spectrum} \\
		\colhead{PTF name} &
		\colhead{$\alpha$(J2000)} & \colhead{$\delta$(J2000)} &
		\colhead{$z$} &
		\colhead{DM\tablenotemark{a}} & 
		\colhead{$A_R$\tablenotemark{b}} & 
		\colhead{$A_r$\tablenotemark{b}} &
		\colhead{UT Date} &
		\colhead{Instrument}  %\\
%& \multicolumn{2}{c}{J2000} &&&&&&& Days
	}
	\startdata
	09cjq\tablenotemark{c} & 21:16:28.483 & $-$00:49:39.71 & 0.02 & 34.69 & 0.16 & 0.17 & 2009-10-22 & Keck-1 --- LRIS  \\
09ecm & 01:06:43.164 & $-$06:22:40.89 & 0.0285 & 35.47 & 0.34 & 0.36 & 2009-10-22 & Keck-1 --- LRIS  \\
09fma & 03:10:23.327 & $-$09:59:58.04 & 0.031 & 35.66 & 0.23 & 0.25 & 2010-01-09 & P200 --- DBSP  \\
10abyy & 05:16:40.524 & $+$06:47:53.76 & 0.0297 & 35.56 & 0.37 & 0.39 & 2011-01-13 & Lick-3m --- Kast  \\
10bgl\tablenotemark{d} & 10:19:04.697 & $+$46:27:23.34 & 0.03 & 35.58 & 0.03 & 0.03 & 2010-02-06 & Keck-1 --- LRIS  \\
10gva & 12:23:55.397 & $+$10:34:50.62 & 0.025 & 35.18 & 0.08 & 0.08 & 2010-06-12 & Keck-1 --- LRIS  \\
10gxi & 12:44:33.681 & $+$31:05:05.35 & 0.0287 & 35.48 & 0.04 & 0.04 & 2010-07-19 & P200 --- DBSP  \\
10jwr & 16:12:15.986 & $+$32:04:14.49 & 0.059 & 37.10 & 0.06 & 0.06 & 2010-07-07 & Keck-1 --- LRIS  \\
10mug & 15:04:06.828 & $+$28:29:17.84 & 0.06 & 37.14 & 0.07 & 0.08 & 2010-08-14 & P200 --- DBSP  \\
10osr & 23:45:45.161 & $+$11:28:42.37 & 0.0235 & 35.04 & 0.11 & 0.12 & 2010-10-11 & Lick-3m --- Kast  \\
10pjg & 23:23:08.010 & $+$13:02:39.18 & 0.0384 & 36.13 & 0.14 & 0.15 & 2010-09-06 & P200 --- DBSP  \\
10qwz & 23:35:18.607 & $+$12:55:31.81 & 0.02 & 34.69 & 0.16 & 0.17 & 2010-10-11 & Lick-3m --- Kast  \\
10rem & 17:17:43.596 & $+$20:52:30.91 & 0.046 & 36.54 & 0.15 & 0.16 & 2010-10-12 & Keck-2 --- DEIMOS  \\
10uls & 01:21:22.659 & $+$04:53:28.75 & 0.044 & 36.44 & 0.08 & 0.09 & 2010-10-11 & Mayall --- RC Spec  \\
10umz & 01:22:01.640 & $-$01:57:23.30 & 0.052 & 36.81 & 0.11 & 0.12 & 2010-10-30 & WHT-4.2m --- ISIS  \\
10uqg & 17:17:00.337 & $+$27:29:27.48 & 0.048 & 36.63 & 0.12 & 0.13 & 2010-10-03 & Keck-1 --- LRIS  \\
10uqn\tablenotemark{e} & 23:06:57.458 & $+$03:56:24.21 & 0.0482 & 36.64 & 0.14 & 0.15 & 2010-10-03 & Keck-1 --- LRIS  \\
10vdl & 23:05:48.879 & $+$03:31:25.54 & 0.016 & 34.19 & 0.15 & 0.17 & 2010-11-07 & Keck-2 --- DEIMOS  \\
10xtq & 08:23:14.292 & $+$21:57:58.00 & 0.08 & 37.79 & 0.11 & 0.12 & 2010-12-06 & P200 --- DBSP  \\
11ajz & 08:26:49.200 & $+$20:22:32.29 & 0.025 & 35.18 & 0.08 & 0.09 & 2011-03-10 & P200 --- DBSP  \\
11cwi & 16:52:28.515 & $+$21:42:00.99 & 0.056 & 36.98 & 0.15 & 0.16 & 2011-05-13 & Mayall --- RC Spec  \\
11go & 11:32:00.235 & $+$53:42:38.06 & 0.0268 & 35.33 & 0.03 & 0.03 & 2011-03-10 & P200 --- DBSP  \\
11hsj & 16:57:58.151 & $+$55:11:01.14 & 0.0287 & 35.48 & 0.05 & 0.05 & 2011-09-29 & Lick-3m --- Kast  \\
11htj & 21:16:03.503 & $+$12:31:20.95 & 0.017 & 34.33 & 0.18 & 0.19 & 2011-10-30 & P200 --- DBSP  \\
11iqb & 00:34:04.836 & $-$09:42:17.92 & 0.0125 & 33.65 & 0.08 & 0.09 & 2011-08-28 & P200 --- DBSP  \\
11izt & 01:52:25.944 & $+$35:30:21.80 & 0.02 & 34.69 & 0.14 & 0.15 & 2011-08-31 & WHT-4.2m --- ISIS  \\
11qax & 23:42:25.584 & $+$00:15:16.83 & 0.022 & 34.90 & 0.06 & 0.07 & 2011-12-18 & Lick-3m --- Kast  \\
12bbm & 11:01:51.229 & $+$45:28:49.60 & 0.0446 & 36.47 & 0.03 & 0.04 & 2012-03-23 & Keck-1 --- LRIS  \\
12bro & 12:24:17.054 & $+$18:55:27.96 & 0.0227 & 34.96 & 0.09 & 0.10 & 2012-04-29 & P200 --- DBSP  \\
12bvh & 10:43:53.752 & $+$11:40:17.89 & 0.0026 & 30.22 & 0.07 & 0.07 & 2012-05-21 & Lick-3m --- Kast  \\
12cod & 13:22:35.288 & $+$54:48:47.11 & 0.0118 & 33.52 & 0.08 & 0.08 & 2012-05-31 & TNG --- DOLORES  \\
12efk & 16:24:43.887 & $+$31:51:37.14 & 0.0931 & 38.14 & 0.05 & 0.06 & 2012-06-16 & Keck-1 --- LRIS  \\
12fip & 15:00:51.041 & $+$09:20:25.12 & 0.034 & 35.86 & 0.07 & 0.08 & 2012-07-21 & P200 --- DBSP  \\
12fo & 12:58:36.924 & $+$27:10:24.94 & 0.026 & 35.26 & 0.04 & 0.04 & 2012-01-27 & Mayall --- RC Spec  \\
12ftc & 15:05:01.880 & $+$20:05:54.63 & 0.0732 & 37.59 & 0.09 & 0.10 & 2012-07-27 & P200 --- DBSP  \\
12gnn & 15:58:49.278 & $+$36:10:10.95 & 0.0308 & 35.64 & 0.06 & 0.07 & 2012-08-21 & WHT-4.2m --- ISIS  \\
12grj & 01:20:39.003 & $+$04:46:23.77 & 0.034 & 35.86 & 0.07 & 0.08 & 2012-07-19 & P200 --- DBSP  \\
12hsx & 00:55:03.328 & $+$42:19:52.01 & 0.019 & 34.57 & 0.23 & 0.25 & 2012-08-19 & Keck-1 --- LRIS  \\
12krf & 22:48:16.673 & $+$24:08:58.25 & 0.0625 & 37.23 & 0.37 & 0.40 & 2012-12-05 & P200 --- DBSP  \\
13aaz & 11:18:56.939 & $+$13:03:50.03 & 0.00269 & 30.30 & 0.06 & 0.07 & 2013-05-02 & P200 --- DBSP  \\
13kg\tablenotemark{e} & 11:34:36.446 & $+$54:53:23.69 & 0.019 & 34.57 & 0.04 & 0.04 & 2013-06-06 & Keck-2 --- DEIMOS  \\
13bjx & 14:14:52.106 & $+$36:47:28.58 & 0.0279 & 35.42 & 0.02 & 0.02 & 2013-08-03 & P200 --- DBSP  \\
13bld\tablenotemark{e} & 16:24:54.586 & $+$41:02:59.24 & 0.0331 & 35.80 & 0.02 & 0.02 & 2013-07-05 & P200 --- DBSP  \\
13bsg & 13:50:07.229 & $+$33:45:07.46 & 0.061 & 37.17 & 0.06 & 0.07 & 2013-07-05 & P200 --- DBSP  \\
13ccu & 02:08:51.874 & $+$41:49:32.80 & 0.074 & 37.61 & 0.19 & 0.21 & 2013-08-14 & P200 --- DBSP  \\
13clj & 01:30:40.412 & $+$14:28:50.04 & 0.056 & 36.98 & 0.11 & 0.12 & 2013-09-04 & P200 --- DBSP  \\
13cly & 00:12:46.944 & $+$04:40:34.59 & 0.0428 & 36.37 & 0.06 & 0.06 & 2013-09-03 & Magellan-Baade --- IMACS  \\
13cnk & 02:02:12.776 & $+$07:58:38.61 & 0.04 & 36.22 & 0.16 & 0.17 & 2013-10-04 & Keck-2 --- DEIMOS  \\
13dkk & 23:41:35.156 & $+$03:43:30.37 & 0.0092 & 32.98 & 0.14 & 0.15 & 2013-10-05 & P200 --- DBSP  \\
13dkz & 01:36:11.577 & $+$33:37:01.43 & 0.016 & 34.19 & 0.11 & 0.12 & 2013-11-02 & P200 --- DBSP  \\
13dla & 01:02:49.095 & $-$00:44:30.80 & 0.0518 & 36.80 & 0.08 & 0.09 & 2013-11-02 & P200 --- DBSP  \\
13dqy & 23:19:44.700 & $+$10:11:04.40 & 0.0119 & 33.54 & 0.10 & 0.11 & 2013-11-26 & P200 --- DBSP  \\
13dzb & 03:10:50.199 & $-$00:21:40.32 & 0.037 & 36.05 & 0.18 & 0.19 & 2013-11-27 & LCOGT --- FLOYDS  \\
14abc & 12:22:57.328 & $+$28:29:54.75 & 0.0254 & 35.21 & 0.06 & 0.06 & 2014-04-04 & P200 --- DBSP  \\
14adz & 13:49:57.814 & $+$37:45:08.49 & 0.078 & 37.73 & 0.03 & 0.03 & 2014-04-29 & Keck-1 --- LRIS  \\
14ajq\tablenotemark{d} & 12:06:32.001 & $+$39:14:13.65 & 0.036 & 35.99 & 0.08 & 0.08 & 2014-04-09 & APO 3.5m --- DIS  \\
14aoi & 12:09:11.543 & $+$29:10:20.90 & 0.012 & 33.56 & 0.05 & 0.05 & 2014-06-30 & Lick-3m --- Kast  \\

\enddata

\tablenotetext{a}{Derived using the \emph{lum\textunderscore dist} routine in MATLAB with the cosmological parameters given in Section \ref{sec:analysis}.} 
\tablenotetext{b}{Derived using the \emph{sky\textunderscore ebv} routine in MATLAB, using $R_V=3.08$.}
\tablenotetext{c}{The spectrum of this SN has a reddened continuum and Na~D absorption lines; it likely suffers from host-galaxy extinction.}
\tablenotetext{d}{The spectrum of this SN has a reddened continuum, but no Na~D absorption lines; it may suffer from host-galaxy extinction.}
\tablenotetext{e}{The spectrum of this SN has a blue continuum, but also Na~D absorption lines; it could possibly suffer from host-galaxy extinction.}
\end{deluxetable*}

\begin{deluxetable*}{llrcrccccc}
	\tablewidth{0pt} 
	\tablecolumns{10}
	\tablecaption{List of derived quantities. \label{tab:SNDerivedQuantities}}
	\tablehead{  % column headings
		\colhead{PTF name} &
		\colhead{$t_0$} &
		\colhead{$t_{\text{rise}}$} &
		\colhead{Peak $R$ Mag} &
		\colhead{$\Delta m_{15}$} & 
		\colhead{$E_{51}/M_{10}$} & 
		\colhead{$v$ H$\alpha$} &
		\colhead{Phase} &
		\colhead{$v_{50}$ H$\alpha$\tablenotemark{a}} &
		\colhead{$M_{\text{Ni}}$} \\
		& \colhead{MJD} & \colhead{days} & \colhead{} & \colhead{} & \colhead{} & \colhead{$10^3$ km s$^{-1}$} & \colhead{days} &\colhead{$10^3$ km s$^{-1}$} & \colhead{M$_\odot$}
	}

	\startdata

	09cjq & $55043.30 \pm 2.96$ & 6.9 & -16.29 & -0.019 & --- & --- & --- & --- & --- \\ 
09ecm & $55083.94 \pm 0.36$\tablenotemark{b} & 12.1 & -17.42 & -0.045 & --- & --- & --- & --- & --- \\ 
09fma & $55108.43 \pm 1.08$ & 5.8 & -17.74 & 0.054 & --- & --- & --- & --- & --- \\ 
10abyy & $55536.29 \pm 0.90$ &  6.9 & -18.75 & 0.35 & $6^{+13}_{-0.83}$ & --- & --- & --- & --- \\ 
10bgl & $55197.62 \pm 1.00$\tablenotemark{b} & 10.3 & -16.77 & 0.14 & --- & --- & --- & --- & --- \\ 
10gva & $55320.28 \pm 0.90$ &  7.3 & -18.26 & 0.067 & $9^{+1.1}_{-6.6}$ & $9.47 \pm 1.1$ & 38 & $8.44 \pm 1.1$ & $0.065 \pm 0.032$ \\ 
10gxi & $55320.87 \pm 2.48$ & 8.4 & -17.19 & 0.013 & --- & --- & --- & --- & --- \\ 
10jwr & $55354.74 \pm 2.49$ & 3.1 & -18.21 & 0.13 & --- & --- & --- & --- & --- \\ 
10mug & $55373.81 \pm 2.49$ & 6.7 & -18.24 & 0.23 & --- & --- & --- & --- & --- \\ 
10osr & $55389.39 \pm 0.93$ & 10.7 & -17.31 & -0.044 & --- & --- & --- & --- & --- \\ 
10pjg & $55385.91 \pm 2.48$ & 7.9 & -16.29 & -0.022 & --- & --- & --- & --- & --- \\ 
10qwz & $55415.87 \pm 1.50$ &  6.5 & -15.83 & -0.024 & $0.73^{+0.056}_{-0.3}$ &  $5.24 \pm 0.99$ & 63 & $5.76 \pm 1.2$ & --- \\ 
10rem & $55415.21 \pm 1.96$ & 3.6 & -16.68 & -0.18 & --- & --- & --- & --- & --- \\ 
10uls & $55445.96 \pm 0.48$ &  9.6 & -17.70 & 0.21 & $1.5^{+3.9}_{-0.21}$ & --- & --- & --- & --- \\ 
10umz & $55444.44 \pm 0.61$\tablenotemark{b} &  14.7 & -17.21 & -0.0021 & $0.82^{+2.4}_{-0.23}$ &  $6.64 \pm 1.6$ & 52 & $6.74 \pm 1.7$ & --- \\ 
10uqg & $55448.65 \pm 1.50$ & 9.1 & -17.97 & 0.41 & --- & --- & --- & --- & --- \\ 
10uqn & $55445.36 \pm 1.88$ & 3.8 & -17.30 & 0.055 & --- & --- & --- & --- & --- \\ 
10vdl & $55452.29 \pm 1.98$ & 5.4 & -15.24 & -0.12 & --- & --- & --- & --- & --- \\ 
10xtq & $55465.99 \pm 0.49$ & 6.5 & -18.42 & 0.21 & --- & --- & --- & --- & --- \\ 
11ajz & $55592.40 \pm 0.96$ &  7.8 & -17.65 & 0.049 & $6.2^{+0.48}_{-4.3}$ &  $9.47 \pm 0.65$ & 37 & $8.34 \pm 0.83$ & --- \\ 
11cwi & $55672.98 \pm 1.48$ & 3.8 & -17.13 & -0.13 & --- & --- & --- & --- & --- \\ 
11go & $55570.91 \pm 1.41$ & 6.4 & -16.37 & 0.089 & --- & --- & --- & --- & --- \\ 
11hsj & $55753.38 \pm 1.95$ & 7.1 & -17.52 & -0.076 & --- & --- & --- & --- & --- \\ 
11htj & $55751.92 \pm 1.48$ & 13.7 & -16.68 & -0.064 & --- & --- & --- & --- & --- \\ 
11iqb & $55764.68 \pm 0.21$\tablenotemark{b} &  8.1 & -18.44 & 0.14 & $9.7^{+0.37}_{-1.9}$ & --- & --- & --- & --- \\ 
11izt & $55765.91 \pm 2.46$ &  7.8 & -15.94 & -0.083 & $0.38^{+0.5}_{-0.23}$ & $7.76 \pm 1.2$ & 37 & $6.88 \pm 1.1$ & $0.027 \pm 0.014$ \\ 
11qax & $55866.69 \pm 0.41$ &  7.3 & -17.23 & -0.023 & $4.8^{+0.77}_{-3.1}$ &  $8.85 \pm 0.65$ & 45 & $8.50 \pm 0.9$ & --- \\ 
12bbm & $55980.38 \pm 0.98$ &  8.2 & -17.40 & 0.17 & $1.4^{+2.3}_{-0.58}$ & --- & --- & --- & --- \\ 
12bro & $56000.85 \pm 0.38$ &  6.8 & -17.22 & 0.015 & $3.9^{+0.3}_{-2.7}$ & $9.12 \pm 1.3$ & 44 & $8.66 \pm 1.4$ & $0.065 \pm 0.033$ \\ 
12bvh & $56002.23 \pm 0.95$ &  7.2 & -16.91 & -0.095 & $2^{+0.15}_{-0.61}$ & $6.48 \pm 0.67$ & 66 & $7.24 \pm 0.96$ & $0.096 \pm 0.048$ \\ 
12cod & $56019.41 \pm 1.90$ &  9.7 & -18.31 & 0.18 & $4^{+18}_{-3}$ & --- & --- & --- & $0.085 \pm 0.042$ \\ 
12efk & $56057.25 \pm 0.71$\tablenotemark{b} &  9.9 & -18.76 & 0.17 & $13^{+3.9}_{-10}$ & --- & --- & --- & --- \\ 
12fip & $56089.23 \pm 0.97$ &  4.9 & -16.60 & -0.023 & $2.1^{+0.33}_{-1.3}$ &  $8.71 \pm 0.89$ & 38 & $7.81 \pm 0.98$ & --- \\ 
12fo & $55927.40 \pm 0.98$ &  2.9 & -16.95 & -0.19 & $3.3^{+0.53}_{-1.7}$ &  $9.57 \pm 2.3$ & 25 & $7.18 \pm 1.8$ & --- \\ 
12ftc & $56090.35 \pm 0.97$ &  5.5 & -17.57 & -0.032 & $5^{+1.2}_{-3.6}$ &  $7.11 \pm 3.7$ & 30 & $5.79 \pm 3.1$ & --- \\ 
12gnn & $56116.39 \pm 0.97$ &  8.7 & -17.64 & 0.18 & $4.8^{+0.57}_{-2.4}$ &  $7.05 \pm 2.2$ & 42 & $6.58 \pm 2.1$ & --- \\ 
12grj & $56123.45 \pm 0.97$ &  5.3 & -16.72 & -0.045 & $2^{+0.15}_{-0.8}$ & --- & --- & --- & $0.027 \pm 0.014$ \\ 
12hsx & $56112.92 \pm 0.25$\tablenotemark{b} &  16.2 & -16.92 & -0.01 & $2.1^{+0.34}_{-1.5}$ & $8.77 \pm 0.58$ & 44 & $8.34 \pm 0.84$ & $0.099 \pm 0.05$ \\ 
12krf & $56234.14 \pm 0.99$ &  9.5 & -18.69 & 0.069 & $16^{+1.9}_{-11}$ & --- & --- & --- & --- \\ 
13aaz & $56371.75 \pm 1.42$ &  9.5 & -14.44 & 0.021 & $0.17^{+0.013}_{-0.056}$ & $4.88 \pm 0.94$ & 42 & $4.54 \pm 0.94$ & $0.0053 \pm 0.0026$ \\ 
13akg & $56389.70 \pm 2.47$ &  8.4 & -15.90 & -0.06 & $0.65^{+0.23}_{-0.5}$ &  $6.03 \pm 0.98$ & 58 & $6.42 \pm 1.2$ & --- \\ 
13bjx & $56442.70 \pm 0.48$ &  5.1 & -17.45 & 0.2 & $5.6^{+0.43}_{-2.4}$ &  $7.93 \pm 0.8$ & 63 & $8.70 \pm 1.1$ & --- \\ 
13bld & $56442.93 \pm 0.46$ & 4.8 & -15.89 & -0.051 & --- & --- & --- & --- & --- \\ 
13bsg & $56451.74 \pm 0.47$ &  5.2 & -17.36 & 0.037 & $5.4^{+0.63}_{-3}$ &  $9.71 \pm 4.7$ & 25 & $7.27 \pm 3.5$ & --- \\ 
13ccu & $56499.38 \pm 0.98$ & 7.1 & -17.96 & 0.26 & --- & --- & --- & --- & --- \\ 
13clj & $56507.90 \pm 0.45$ &  10.4 & -18.26 & 0.32 & $10^{+0.77}_{-4.1}$ & --- & --- & --- & --- \\ 
13cly & $56505.84 \pm 0.15$\tablenotemark{b} &  8.5 & -17.70 & 0.073 & $4.6^{+0.54}_{-2.1}$ &  $8.20 \pm 1.1$ & 62 & $8.94 \pm 1.4$ & --- \\ 
13cnk & $56509.90 \pm 0.46$ &  8.7 & -16.27 & 0.14 & $1.5^{+0.17}_{-0.63}$ &  $3.94 \pm 0.25$ & 57 & $4.16 \pm 0.43$ & --- \\ 
13dkk & $56546.84 \pm 0.35$ &  5.3 & -14.60 & -0.069 & $0.23^{+0.018}_{-0.13}$ &  $5.46 \pm 0.51$ & 23 & $3.96 \pm 0.44$ & --- \\ 
13dkz & $56547.93 \pm 0.45$ &  5.5 & -16.29 & -0.077 & $1.4^{+0.11}_{-0.61}$ &  $8.00 \pm 1.1$ & 49 & $7.95 \pm 1.2$ & --- \\ 
13dla & $56548.95 \pm 0.45$ &  8.8 & -18.25 & 0.22 & $11^{+1.3}_{-6.6}$ &  $10.74 \pm 1.9$ & 47 & $10.43 \pm 2$ & --- \\ 
13dqy & $56570.79 \pm 0.45$ &  6.8 & -17.59 & 0.24 & $5.4^{+0.41}_{-1.7}$ &  $7.76 \pm 0.44$ & 51 & $7.79 \pm 0.75$ & --- \\ 
13dzb & $56602.81 \pm 0.45$ &  6.2 & -17.24 & 0.038 & $4^{+0.31}_{-1.5}$ &  $10.29 \pm 1.8$ & 19 & $6.98 \pm 1.3$ & --- \\ 
14abc & $56732.28 \pm 2.92$ & 3.8 & -17.25 & 0.086 & --- & --- & --- & --- & --- \\ 
14adz & $56735.34 \pm 0.36$\tablenotemark{b} & 13.0 & -18.36 & 0.1 & --- & --- & --- & --- & --- \\ 
14ajq & $56743.40 \pm 1.99$ & 15.3 & -16.39 & 0.57 & --- & --- & --- & --- & --- \\ 
14aoi & $56769.14 \pm 0.05$\tablenotemark{b} & 6.6 & -15.66 & 0.13 & --- & --- & --- & --- & --- \\

\enddata

\tablenotetext{a}{Estimated using the relation from \cite{faran_photometric_2014}; see Section \ref{sec:analysis}.}
\tablenotetext{b}{From exponential fit.}

\end{deluxetable*}

\begin{deluxetable*}{lccccl}
	\tabletypesize{\footnotesize}
	\tablecolumns{6}
	\tablewidth{0pt}
	\tablecaption{Explosion parameters from events in the literature. \label{tab:Ni56MassLit}}
	\tablehead{
		\colhead{SN} &
		\colhead{$E_{51}/M_{10}$} &
		\colhead{\Ni{}/M$_\odot$} &
		\colhead{$E_{51}/M_{10}$\tablenotemark{a}} &
		\colhead{\Ni{}/M$_\odot$\tablenotemark{a}} &
		\colhead{Reference} }

		\startdata
		SN 2004et & 0.63 & 0.06 & --- & --- & \cite{maguire_optical_2010} \\
		SN 2005cs & 0.29 & 0.003 & $0.29_{-0.2}^{+0.05}$ & $0.0034 \pm 0.0017$ & \cite{pastorello_sn_2009} \\
		SN 2007od & 0.8 & 0.02 & --- & ---& \cite{inserra_quantitative_2012} \\
		SN 2009E & 0.32 & 0.04 & --- & --- & \cite{pastorello_sn_2012}\\
		SN 2009N & 0.42 & 0.02 & --- & ---& \cite{takats_sn_2014} \\
		SN 2012A & 0.38 & 0.011 & --- & ---& \cite{tomasella_comparison_2013} \\
		SN 2012aw & 1.1 & 0.056 & $1.41_{-1.09}^{+0.13}$ & $0.074 \pm 0.037$ & \cite{dallora_type_2014} \\
		SN 2012ec & 0.95 & 0.04 & --- & --- & \cite{barbarino_sn_2015} \\
		SN 2013ab & 0.5 & 0.06 & $2.96_{-2.12}^{+0.27}$ & $0.072 \pm 0.036$ & \cite{bose_sn_2015} 
	\enddata
	\tablenotetext{a}{This analysis.}
\end{deluxetable*}

\clearpage
\appendix

\section{Light-Curve Parameter Estimation}
\label{sec:lightCurveParamEstimation}

\subsection{Light-Curve Smoothing Algorithm}

The smoothing was performed using a smoothing kernel of the following functional form:
\begin{equation}
	K(t,\tau) = N(t,\tau,\sigma(\tau)),
\end{equation}

\n where $N$ is a normal distribution evaluated at time $t$ with mean $\tau$ and standard deviation $\sigma(\tau)$ defined by
\begin{equation}
	\sigma (\tau) = \begin{cases} 
		1, & \tau \leq 5 \\
		10, & \tau \geq 50 \\
		0.2\tau, & \text{else,}
	\end{cases}
\end{equation}

\n where $\tau$ is measured in days from explosion. For each time $\tau$, we fit a straight line by solving the least-squares problem
\begin{equation}
	f_i = \begin{pmatrix} t_i & 1 \end{pmatrix} \begin{pmatrix} a(\tau) \\ b(\tau) \end{pmatrix} 
\end{equation}

\n with weights
\begin{equation}
	w_i = K(t_i, \tau, \sigma(\tau))/e_i^2,
\end{equation}

\n where $f_i$ is the flux with error $e_i$ at time $t_i$. This method has the advantage that it is adaptive to the physically different time scales of the light curve. During the rise, the light curve changes on a short time scale ($<1$ day), while during the plateau the time scale is longer (1--2 weeks). We used linear interpolation to fill gaps in the data of greater than 20 days. In addition, we occasionally added auxiliary data points when the smoothed function deviated wildly from a reasonable fit. The resulting smoothed light curves are shown in Figures \ref{fig:smoothedLC1}---\ref{fig:smoothedLC4}.

\subsection{$t_0$ from Exponential Fits}
Using a similar parametrization to that of \cite{ofek_interaction-powered_2014}, we used nonlinear least squares to fit an exponential to the early-time data:
\begin{equation}
	f(t) = f_m \left(1 - \exp\left(-\frac{t-t_0}{t_e}\right)\right),
\end{equation} 

\n where $t$ is the time in days, $f_m$ is the peak flux, $t_0$ is the time of explosion, and $t_e$ is the characteristic rise time. The resulting fits are shown in Figure \ref{fig:expFits}. The uncertainties in the parameters were estimated using the 95\% confidence levels.

\section{Extended Data}

% Light curve figures
\begin{figure*}[ht]
	\centering
	\input{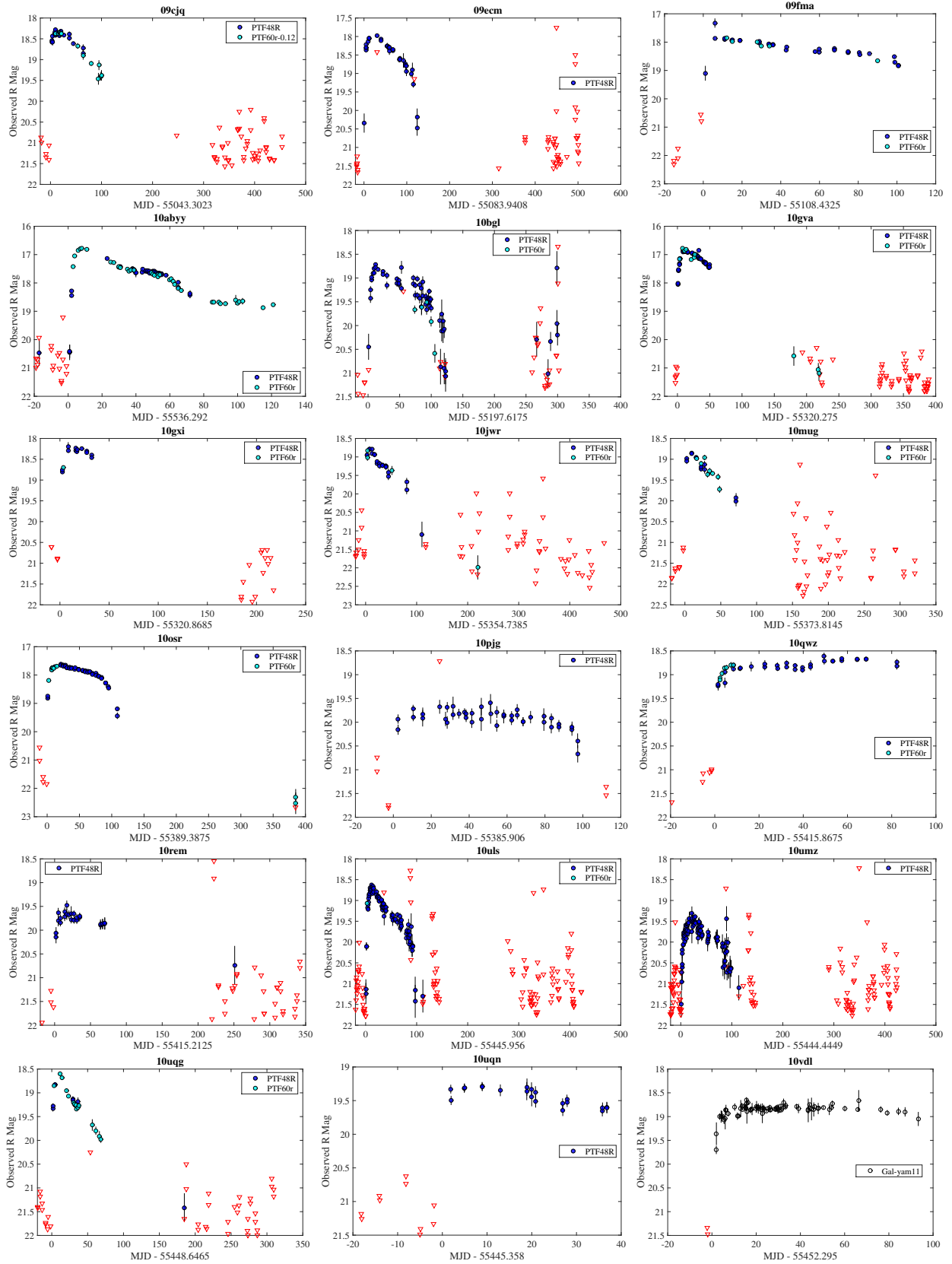}
	\caption{$R$-band light curves. Inverted red triangles represent upper limits. Note that as discussed in the text (Section \ref{sec:analysis}), several have been supplemented with data taken either from the literature (PTF12bvh, iPTF13aaz, PTF10vdl) or from a different telescope (PTF12cod, iPTF13dkk, iPTF13dzb, iPTF13dqy). We have found that small additive constants (indicated in the figures) are needed to make the supplementary data consistent with the PTF observations; this is due most likely to slightly different filter responses.}
	\label{fig:lcPlots1}
\end{figure*}
\begin{figure*}[ht]
	\centering
	\input{lcPlotsLatex2.txt}
	\caption{Same as Figure \ref{fig:lcPlots1}.}
	\label{fig:lcPlots2}
\end{figure*}
\begin{figure*}[ht]
	\centering
	\input{lcPlotsLatex3.txt}
	\caption{Same as Figure \ref{fig:lcPlots1}.}
	\label{fig:lcPlots3}
\end{figure*}
\begin{figure*}[ht]
	\centering
	\input{lcPlotsLatex4.txt}
	\caption{Same as Figure \ref{fig:lcPlots1}.}
	\label{fig:lcPlots4}
\end{figure*}

%Exponential fits for t0
\begin{figure}
	\centering
	\input{expFitsLatex.txt}
	\caption{Exponential fits to selected light curves, where limits did not give satisfactory constraints. Red markers were excluded from the fit.}
	\label{fig:expFits}
\end{figure}

\begin{figure}
	\centering
	\input{smoothedLCs1.txt}
	\caption{Smoothed $R$-band light curves. Shown are the data (blue points), auxiliary points defined to improve the interpolation (green squares), and the final smoothed light curve (red solid line).}
	\label{fig:smoothedLC1}
\end{figure}
\begin{figure}
	\centering
	\input{smoothedLCs2.txt}
	\caption{Same as Figure \ref{fig:smoothedLC1}.}
	\label{fig:smoothedLC2}
\end{figure}
\begin{figure}
	\centering
	\input{smoothedLCs3.txt}
	\caption{Same as Figure \ref{fig:smoothedLC1}.}
	\label{fig:smoothedLC3}
\end{figure}
\begin{figure}
	\centering
	\input{smoothedLCs4.txt}
	\caption{Same as Figure \ref{fig:smoothedLC1}.}
	\label{fig:smoothedLC4}
\end{figure}

\clearpage

\begin{figure}
	\centering
	\input{rwGridLatex1.txt}
	\caption{$\chi^2$ of RW11 models for various progenitor radii and $E_{51}/M_{10}$ along the profile of best-fit time of explosion. The white, blue, and teal regions represent the 68\%, 95\%, and 99.7\% confidence regions, respectively.}
	\label{fig:RWGridDensity1}
\end{figure}

\begin{figure}
	\centering
	\input{rwGridLatex2.txt}
	\caption{Same as Figure \ref{fig:RWGridDensity1}.}
	\label{fig:RWGridDensity2}
\end{figure}

\begin{figure}
	\centering
	\input{rwGridLatex3.txt}
	\caption{Same as Figure \ref{fig:RWGridDensity1}.}
	\label{fig:RWGridDensity3}
\end{figure}

\begin{figure}
	\centering
	\input{rwfitsLatex1.txt}
	\caption{Best-fit RW11 models to the data. The best-fit radius $R_*$, energy per unit mass $E_{51}/M_{10}$, and error scaling factor CE are shown in each figure. Filled symbols are the points that were included in the fit.}
	\label{fig:RWFits1}
\end{figure}

\begin{figure}
	\centering
	\input{rwfitsLatex2.txt}
	\caption{Same as Figure \ref{fig:RWFits1}.}
	\label{fig:RWFits2}
\end{figure}

\begin{figure}
	\centering
	\input{rwfitsLatex3.txt}
	\caption{Same as figure \ref{fig:RWFits1}.}
	\label{fig:RWFits3}
\end{figure}

\begin{figure}
	\centering
	\input{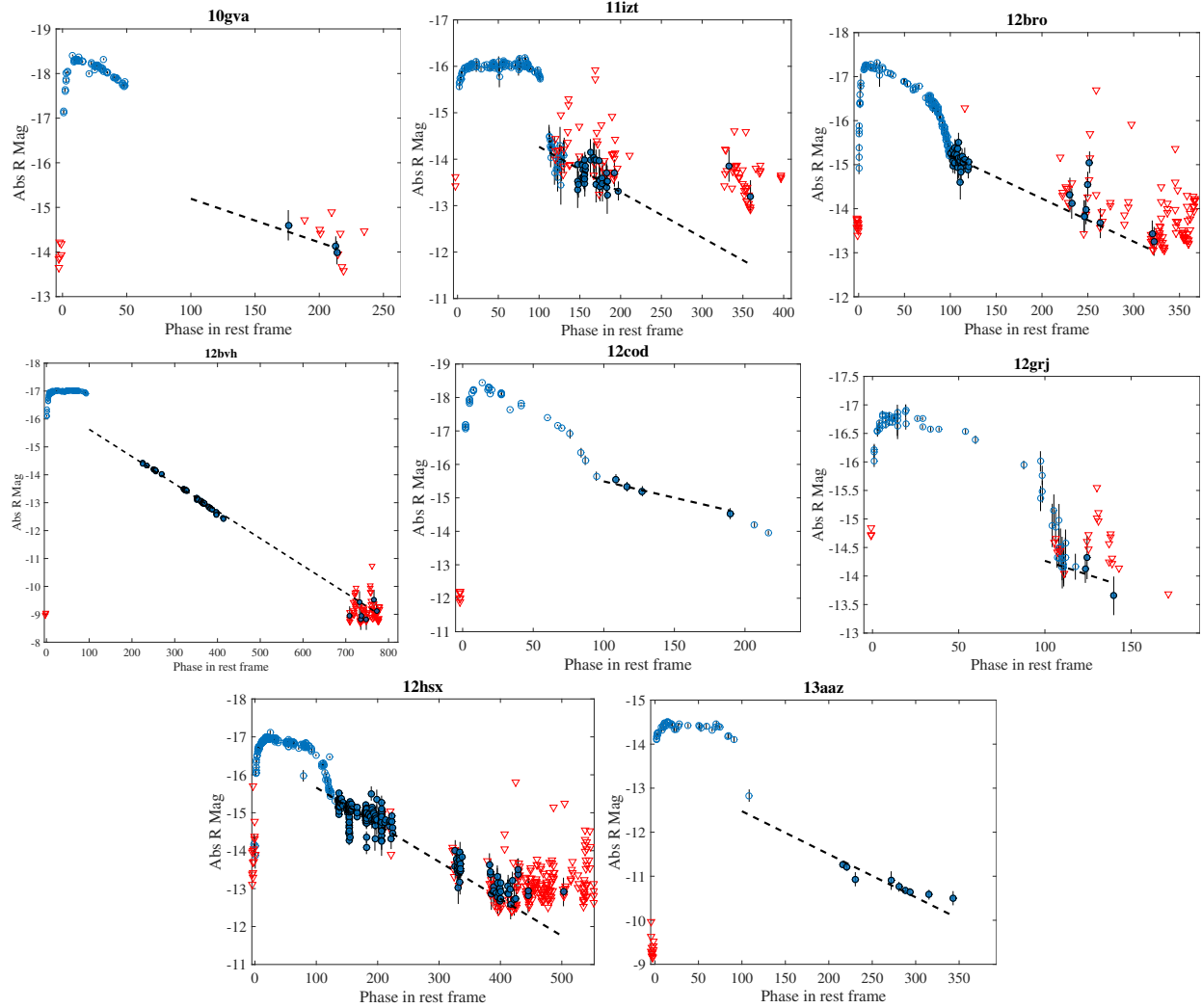}
	\caption{Fits to determine $^{56}$Ni mass. Filled symbols have been included in the fit. The dashed line represents the best fit.}
	\label{fig:ni56Fits}
\end{figure}

\end{document}